\documentclass[twocolumn,structabstract]{aa}   
\usepackage{graphicx}
\usepackage{amsmath,amssymb,amsfonts}
\usepackage{txfonts}
\usepackage{natbib}
\bibpunct{(}{)}{;}{a}{}{,} % to follow the A&A style
\providecommand{\eto}{\ensuremath{\mathrm{e}^}}               %euler exponentials
\providecommand{\iu}{\ensuremath{\mathrm{i}}}                 %imaginary unit
\providecommand{\diff}[1]{\ensuremath{\mathrm{d}#1\,}}        %differential operator
\providecommand{\fdiff}[1]{\ensuremath{\mathrm{d}#1}}         %differential operator alone in fractions, no spacing
\providecommand{\ddiff}[2]{\ensuremath{\mathrm{d}^{#2}#1\,}}  %differential operator with exponent
\providecommand{\deriv}[3][]{\ensuremath{\frac{\mathrm{d}^{#1}#2}{\mathrm{d} #3^{#1}}}} %derivatives, use \deriv{f(x)}{x} or \deriv[n]{f(x)}{x}
\providecommand{\pderiv}[3][]{\ensuremath{\frac{\partial^{#1}#2}{\partial #3^{#1}}}} %partial derivatives, use \pderiv{f(x)}{x} or \pderiv[n]{f(x)}{x}
\begin{document}

\title{Constrained probability distributions of correlation functions}

\author{D.~Keitel\inst{\ref{inst1}, \ref{inst2}}\and P.~Schneider\inst{\ref{inst1}}}
\institute{Argelander-Institut f\"ur Astronomie, Universit\"at Bonn, Auf dem H\"ugel 71, 53121 Bonn, Germany \\ \email{peter@astro.uni-bonn.de}\label{inst1}
\and Max-Planck-Institut f\"ur Gravitationsphysik, Callinstra{\ss}e 38, 30167 Hannover, Germany \\ \email{david.keitel@aei.mpg.de}\label{inst2}
}

% \date{Received / Accepted }

\abstract
% context
{Two-point correlation functions are used throughout cosmology as a measure for the statistics of random fields. When used in Bayesian parameter estimation, their likelihood function is usually replaced by a Gaussian approximation. However, this has been shown to be insufficient.}
% aims
{For the case of Gaussian random fields, we search for an exact probability distribution of correlation functions, which could improve the accuracy of future data analyses.}
% methods
{We use a fully analytic approach, first expanding the random field in its Fourier modes, and then calculating the characteristic function. Finally, we derive the probability distribution function using integration by residues. We use a numerical implementation of the full analytic formula to discuss the behaviour of this function.}
% results
{We derive the univariate and bivariate probability distribution function of the correlation functions of a Gaussian random field, and outline how higher joint distributions could be calculated. We give the results in the form of mode expansions, but in one special case we also find a closed-form expression. We calculate the moments of the distribution and, in the univariate case, we discuss the Edgeworth expansion approximation. We also comment on the difficulties in a fast and exact numerical implementation of our results, and on possible future applications.}
{}


\maketitle

\authorrunning{D. Keitel \& P. Schneider}

\section{Introduction}
 \label{sec:intro}
 
 In several fields of science, there are observations that can be modelled as random processes, either time series or spatial random fields. In cosmology, mostly random fields are of interest, for example as the density perturbation field. Therefore, we use the language of random fields in this article. Still, all results are applicable to time series as well.

 An important statistical quantity of random fields is the two-point correlation function, labelled $\xi(x)$ in the following, with $x$ being the separation between two points of the field. When the empirical correlation function has been measured, it can be used to estimate the parameters of a theoretical model for the random field. The standard method for this, at least in cosmology, is Bayesian inference, using Bayes' theorem:
 \begin{equation}
  \label{eq:bayes_theorem}
  p(\theta|\xi) = \frac{p(\xi|\theta) p(\theta)}{\int \diff{\theta'} p(\xi|\theta') p(\theta')} \, ,
 \end{equation}
 where $\theta$ are the model parameters, $p(\theta|\xi)$ is the posterior probability of some parameters given the data $\xi$, $p(\xi|\theta)$ is called the likelihood and $p(\theta)$ is the prior probability of the parameters. The prior can be chosen, more or less freely, by several schemes, and the integral in the denominator is usually not relevant for parameter estimation studies, as only posterior ratios are necessary. But it is crucial to the Bayesian analysis to determine the correct likelihood function for the problem beforehand, a process also known as forward modelling.

 However, in many applications, such as those most relevant for cosmology, it is very difficult to obtain the correct likelihood function either empirically or theoretically. Therefore, in those cases where the underlying random field is assumed to be Gaussian, it has been standard practice for some time to simply use a Gaussian approximation for the likelihood, as well. Examples include \cite{Fu2007} in a cosmic shear analysis, \cite{Okumura2007} for luminous red galaxy counts, and \cite{Seljak1993} for the cosmic microwave background correlation function.

 There is no a priori reason to assume that the Gaussian approximation should be exact, or even very accurate. In fact, it was shown by \cite{Hartlap2009} that there is a significant deviation between the real likelihood and a Gaussian for simulated cosmic shear data. In this case, the error bars on cosmological parameters improved by 10-40\% when using a non-Gaussian likelihood. This result encouraged a mathematical study on the properties of correlation functions of Gaussian random fields. \cite{Schneider2009} found that these correlation functions cannot take arbitrary values, but are subject to constraints. This can be seen from first studying the power spectrum of the field, which is Fourier conjugate to the correlation function. Power spectra are always non-negative, $P(\vec{k}) \geq 0$ for all wave vectors $\vec{k}$.

 If the correlation function is measured over a separation vector (or `lag parameter') $\vec{x}$ and its integer multiples $n\vec{x}, \ n \in \mathbb{N}$, the non-negativity of $P(\vec{k})$ leads to constraints in the form of inequalities $r_{n\mathrm{l}} \leq r_n \leq r_{n\mathrm{u}}$. Here, $r_n=\xi(n \vec{x})/\xi(0)$ is the `correlation coefficient' normalised by the correlation at zero separation, and the upper and lower boundaries $r_{n\mathrm{l}}$ and $r_{n\mathrm{u}}$ are functions of all $r_i$ with $i<n$. The three lowest order constraints are $0 \leq \xi(0)$, $-1 \leq r_1 \leq 1$ and $-1+2r_1^2 \leq r_2 \leq 1$.

 Since a Gaussian or multivariate Gaussian is unbounded, the existence of the constraints already demonstrates that the real likelihood function cannot be exactly Gaussian, but at least must have its tails cut off. Still, the constraints do not immediately lead to a full description of the likelihood function. Another study by \cite{Sato2011} has used the copula approach to construct a more realistic likelihood function for large-scale structure data. However, this approach is mostly useful for numerical application and does not yield direct insights into the analytical structure of the likelihood function. Also, a preliminary investigation by \cite{Wilking2011} has shown that a simple approach using a Gaussian copula cannot correctly reproduce the likelihood under constraints. Therefore, in this article we will focus on a simple type of random fields, but try to find a fully analytical expression for the likelihood function, or probability distribution, of correlation functions, and will demonstrate that it is indeed manifestly non-Gaussian.

 For this derivation, we concentrate on Gaussian random fields, since their unique properties allow for a fully analytical calculation. In addition, a Gaussian field is fully specified by its two-point statistics, i.e. $\xi(\vec{x})$ or equivalently $P(\vec{k})$, which makes the derivation of their probability distributions especially rewarding, since then $P(\vec{k})$ determines the full set of joint probability distributions $p(\xi_1)$, $p(\xi_1,\xi_2)$, $p(\xi_1,\xi_2,\xi_3)$ and so on, with $\xi_i=\xi(\vec{x}_i)$.

 This article consists of five main sections, apart from this introduction. In Sect. \ref{sec:univar}, we derive the univariate probability distribution function. We also calculate its moments and present explicit results for a special power spectrum. Then, we repeat the derivation for bivariate distributions in Sect. \ref{sec:bivar}, also discussing its moments. We go on to discuss possible numerical implementations of these results in Sect. \ref{sec:numerical}. Using numerical evaluation, we can discuss the properties of the distribution functions in more detail in Sect. \ref{sec:discuss}. There, we comment on the general analytical properties of the uni- and bivariate functions. We also use the moments to construct an Edgeworth expansion of the univariate distribution, and we generalise our derivations and results to higher dimensions. We conclude the article in Sect. \ref{sec:conclusions}.

\section{Univariate distribution}
 \label{sec:univar}
 
 \subsection{Derivation}
  \label{sec:univar_derivation}
  We describe a real Gaussian random field by its Fourier decomposition
  \begin{equation}
   g(\vec{x})=\sum \limits_{n=-\infty}^{\infty}g_n \eto{\iu \vec{k}_n \vec{x}} \, .
  \end{equation}
  The Fourier modes are independently distributed, each with a Gaussian probability distribution:  
  \begin{equation}
   p(g_n)=\frac{1}{\pi \sigma_n^2} \eto{-|g_n|^2 / \sigma_n^2} \, .
  \end{equation}
  The mode dispersions are determined by the power spectrum,
  \begin{equation}
   \label{eq:derivation_sigma}
   \sigma_n^2 = \frac{1}{L^{N_{\mathrm{dim}}}}P(|\vec{k}_n|) \, .
  \end{equation}
  The following derivations will be independent of the choice of a power spectrum, as long as it obeys the constraint of non-negativity. So the $\sigma_n$ are arbitrary parameters.
  
  As is usually done in numerical simulations, we consider a finite field, $\vec{x} \in [0,L]^{N_{\mathrm{dim}}}$, with periodic boundary conditions. A sufficiently large finite field will be representative of a field on the whole of $\mathbb{R}^{N_{\mathrm{dim}}}$ if the random field has no power on scales larger than $L$. This is also equivalent to the assumption of statistical homogeneity on these scales. Together with isotropy, we know that the correlation function depends on the distance modulus, or lag parameter, only:
  \begin{equation}
   \xi(\vec{x},\vec{y})=\xi(|\vec{x}-\vec{y}|) \, .
  \end{equation}
  We will now concentrate on a one-dimensional field to keep expressions simple. However, in Sect. \ref{sec:univar_ndim} we will see that all results also hold for higher dimensions. We start with an estimator for the correlation function from the finite field, given by
  \begin{equation}
   \label{eq:pxi_derivation_xiestimator}
   \xi(x) = \langle g(y)g^*(x+y) \rangle = \frac{1}{L} \int\limits_0^L \diff{y} g(y)g^*(x+y) \, ,
  \end{equation}
  which approaches the true value for $L \rightarrow \infty$.
  
  Going to $k$-space, the Fourier modes that fit inside the interval $[0,L]$ are discrete. Each wave number $k_n$ has to fulfil the condition $\eto{\iu k_n L} = 1$, so that
  \begin{equation}
   \label{eq:pxi_derivation_kn}
   k_n = \frac{2 \pi}{L} n
  \end{equation}
  with $n \in \mathbb{N}$. For a real-valued random field, the Fourier components fulfil $g_{-n}=g_n^*$. Using this property, the mode expansion of the field can be split up as
  \begin{equation}
   \label{eq:pxi_derivation_modeexpansion}
   g(x) = \sum\limits_{n=-\infty}^{\infty} g_n \eto{\iu k_n x} = \sum\limits_{n=1}^{\infty} \left( g_n \eto{\iu k_n x} + g_n^* \eto{-\iu k_n x} \right) + g_0 \, .
  \end{equation}
  Without loss of generality, we assume that the field has zero mean, since we can always achieve this by a simple transformation. Then, the zero mode $g_0$ cancels out. We can then insert his expansion into the estimator (\ref{eq:pxi_derivation_xiestimator}). For the spatial integrals, we can use the integral representation of the \emph{Kronecker delta symbol}:
  \begin{equation}
   \int\limits_{0}^{L} \diff{x} \eto{ \iu \left(2\pi / L \right) x ( n-m )} = L \delta_{nm} = \begin{cases}
                   L & \text{if} \quad n =    m \, . \\
                   0 & \text{if} \quad n \neq m \, .
                 \end{cases}
  \end{equation}
  The correlation function is then given by
  \begin{align}
   \xi(x) = \frac{1}{L} \sum\limits_{n=1}^{\infty} \sum\limits_{m=1}^{\infty} L & \left(
                      \ g_n   g_m   \eto{ \iu k_m x} \delta_{ n,-m}
                      + g_n^* g_m   \eto{ \iu k_m x} \delta_{ n, m} \right. \nonumber \\
             & \left. + g_n   g_m^* \eto{-\iu k_m x} \delta_{ n, m}
                      + g_n^* g_m^* \eto{-\iu k_m x} \delta_{-n, m} \right) \, .
  \end{align}
  Executing the sum over $m$, only half of the terms survive, and the remaining exponentials give a cosine function:
  \begin{equation}
   \label{eq:univar_derivation_xi}
   \xi(x) = 2 \sum\limits_{n=1}^{\infty} |g_n|^2 \cos ( k_n x ) \, .
  \end{equation}
  Now that we have a convenient expression for (the estimator of) the correlation function, we need to take one more intermediate step before calculating its probability distribution. This is the characteristic function, which, in general, is defined as the Fourier transform of a probability distribution function. For the given random field, we can calculate the characteristic function by means of an ensemble average:
  \begin{equation}
   \psi(s)
           = \left\langle \eto{\iu s \xi(x)} \right\rangle_{x}
           = \left( \prod\limits_{n=1}^{\infty} \int \ddiff{g_n}{2} p(g_n) \right) \eto{\iu s \xi(x)} \, .
  \end{equation}
  Since the field is Gaussian, the modes are independently distributed, and the probability distribution factorises. Inserting (\ref{eq:univar_derivation_xi}), we get
  \begin{equation}
   \psi(s)
          =  \prod\limits_{n=1}^{\infty} \int \ddiff{g_n}{2} p(g_n) \, \eto{2 \iu s |g_n|^2 \cos(k_nx)}
          =: \prod\limits_{n=1}^{\infty} \psi_n(s) \, .
  \end{equation}
  In the individual factors $\psi_n(s)$, we substitute $z = |g_n|^2$ to solve the integral:
  \begin{align}
   \psi_n(s)
             &= \int\limits_0^{2\pi} \diff{\phi_n} \int\limits_0^{\infty}  \frac{\diff{|g_n|} |g_n|}{\pi \sigma_n^2} \exp\left(-\frac{|g_n|^2}{\sigma_n^2}\right) \exp\left(2 \iu s |g_n|^2 \cos (k_nx) \right) \nonumber \\
             &= \frac{1}{\sigma_n^2} \int\limits_0^{\infty} \diff{z} \exp\left(-z \frac{1 - 2 \iu s \sigma_n^2 \cos (k_nx)}{\sigma_n^2}\right) \nonumber \\
             &= \frac{1}{1 - 2 \iu s \sigma_n^2 \cos (k_nx)} \, .
  \end{align}
  With the product over all modes, we obtain the full characteristic function as
  \begin{equation}
   \label{eq:derivation_pxi_charfct}
   \psi(s) = \prod\limits_{n=1}^{\infty} \frac{1}{1 - 2 \iu s \sigma_n^2 \cos (k_nx)} = \prod\limits_{n=1}^{\infty} \frac{1}{1 - 2 \iu s C_n} \, ,
  \end{equation}
  where in the last step we introduced the shorthand notation
  \begin{equation}
   \label{eq:derivation_cnfactors}
   C_n = \sigma_n^2 \cos(k_nx) \, .
  \end{equation}
  The characteristic function is an important result in its own right, since we can use it to calculate the moments of the distribution, which we will do in Sect. \ref{sec:univar_moments}. But for now, we go on to calculate the probability density distribution $p(\xi)$ by an inverse Fourier transform:
  \begin{equation}
   p(\xi) = \int\limits_{-\infty}^{\infty} \frac{\fdiff{s}}{2\pi} \eto{-\iu s \xi} \psi(s)
          = \int\limits_{-\infty}^{\infty} \frac{\fdiff{s}}{2\pi} \eto{-\iu s \xi} \prod\limits_{n=1}^{\infty} \frac{1}{1 - 2 \iu s C_n} \, .
  \end{equation}
  We will solve this integral using the theorem of residues, since the integrand is analytic except at its poles
  \begin{equation}
   \label{eq:derivation_poles}
   s_n = \frac{-\iu}{2 C_n} = \frac{-\iu}{2 \sigma_n^2 \cos (k_nx)} \, .
  \end{equation}
  All of these lie on the imaginary axis. However, if $\cos (k_nx)=0$ for some $n$, then the corresponding factor in the characteristic function is unity, and there is no pole and no contribution to the integral from this term. On the other hand, some $C_n$ may be equal, and so there may be poles of higher order (multiple poles). These special cases will be discussed in appendix \ref{sec:multipoles}, but for now we focus on the standard case of $N$ simple poles.
   
  We can choose a contour of integration made up of two parts, a straight section $[R,R]$ combined with a semicircle of radius $R$, which we parametrise by $s=R\eto{\iu \phi}$, with either $\phi \in [0,\pi]$ for the upper or $\phi \in [\pi,2\pi]$ for the lower half-plane. To get the full integral for $p(\xi)$, we have to take the limit of $R \rightarrow \infty$. The numerator of the integrand is $\eto{-\iu s \xi}$, whereas the denominator, after executing the product, is only a polynomial in $s$. So the numerator dominates the convergence behaviour, requiring $\Re(-\iu s \xi) < 0$. For $\xi>0$, this corresponds to $\Im(s) < 0$, and we can close the contour in the lower half-plane. If instead $\xi<0$, the requirement is $\Im(s) > 0$, and we can close the contour in the upper half-plane. So for a given $\xi$, only the poles lying in the corresponding half-plane contribute to the sum of residues. We encode this behaviour in the factor
  \begin{equation}
   \mathcal{H}_n = H(\xi)H(C_n) - H(-\xi)H(-C_n) \, ,
  \end{equation}
  where for the Heaviside step function we use the convention
  \begin{equation}
    H(x) = \begin{cases}
            1   & \text{if} \quad x > 0 \, . \\
	    0.5 & \text{if} \quad x = 0 \, . \\
            0   & \text{if} \quad x < 0 \, .
           \end{cases}
   \end{equation}
  If all poles $s_n$ are simple, we can calculate the residues by
  \begin{align}
   \operatorname{Res}_{s_n} &= \lim\limits_{s \rightarrow s_n} \left( (s-s_n) \frac{\eto{-\iu s \xi}}{1-2 \iu s C_n} \prod\limits_{m \neq n} \frac{1}{1-2 \iu s C_m} \right) \nonumber \\
                            &= \eto{-\xi/(2C_n)} \frac{\iu}{2C_n} \prod\limits_{m \neq n} \frac{1}{1-\frac{C_m}{C_n}} \, .
  \end{align}
  Inserting the winding numbers $w_n=1$ for the upper and $w_n=-1$ for the lower contour, the full integral is then
  \begin{align}
   p(\xi) &= \int\limits_{-\infty}^{\infty} \frac{\fdiff{s}}{2\pi} \eto{-\iu s \xi} \prod\limits_{n=1}^{\infty} \frac{1}{1 - 2 \iu s C_n}
          = 2 \pi \iu \sum\limits_n w_n \operatorname{Res}_{s_n} \nonumber \\
          &= \sum\limits_{n=1}^{\infty}  \mathcal{H}_n \eto{-\xi/(2C_n)} \frac{1}{2C_n} \prod\limits_{m \neq n} \frac{1}{1-\frac{C_m}{C_n}} \, .\label{eq:univar_derivation_pxifinal}
  \end{align}
  This result holds for most relevant combinations of input power spectra and lag parameters. For other cases, we derive a generalised result of a very similar functional form, that also holds for multiple poles, in appendix \ref{sec:multipoles}. 
  
  We were unable to further simplify the limit of the infinite sum in the probability distribution function (\ref{eq:univar_derivation_pxifinal}). However, as long as the power spectrum decreases at least like $k^{-2}$ for large $k$, our numerical implementation of the sum formulae, as described in Sect. \ref{sec:numerical}, showed that the probability distribution function converges as well. In practice, it is therefore possible to truncate the series at some maximum mode number $N$ without losing much precision.

  Also, it is obvious from  Eq. (\ref{eq:univar_derivation_pxifinal}) that for large $\xi$, a single mode will always dominate the sum, so that asymptotically, the distribution is not Gaussian, $p(\xi) \propto \eto{-\xi^2}$, but instead exponential, $p(\xi) \propto \eto{-\xi / \left(2C_{\mathrm{max}}\right)}$.

  We also note at this point that Eq. (\ref{eq:univar_derivation_pxifinal}) depends on the field size $L$, separation $x$ and power spectrum $P(|k_n|)$ only through the ratios $x/L$ and $P(|k_n|)/L$, as can be seen from the definition of $C_n$ in Eq. (\ref{eq:derivation_cnfactors}). When we present numerical results in the further course of this article, we therefore give these quantities only. Furthermore, in the case of a Gaussian power spectrum,
  \begin{equation}
   \label{eq:pxi_derivation_gausspwrspec}
   P(|k_n|) = \frac{1}{\sigma_P \sqrt{2\pi}} \eto{-k_n^2/\left(2\sigma_P^2\right)} \, ,
  \end{equation}
  we can directly state $L \, \sigma_P$ as the relevant quantity.
 
\begin{figure}
   \resizebox{\hsize}{!}{\includegraphics[angle=270]{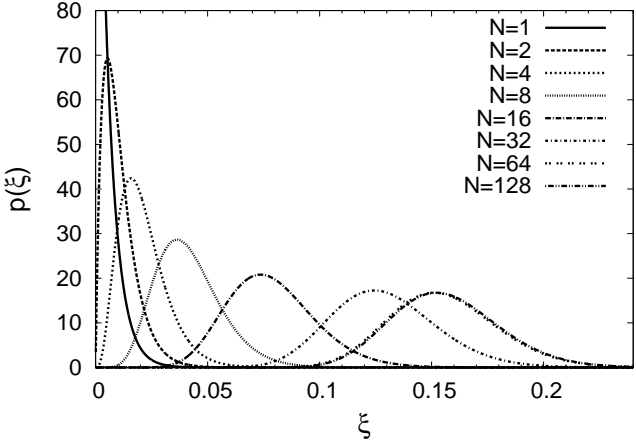}}
   \caption{\small Univariate distributions for different mode numbers $N$, demonstrating convergence. All distributions have a Gaussian power spectrum with $L \, \sigma_P=150$ and lag $x=0$. With maxima from left to right: \newline \textit{solid}: $N=1$, \textit{long-dashed}: $N=2$, \textit{dashed}: $N=4$, \textit{dotted}: $N=8$, \textit{long-dashed-dotted}: $N=16$, \textit{dashed-dotted}: $N=32$, \textit{double-dashed}: $N=64$, \textit{long-dashed-double-dotted}: $N=128$.}
   \label{fig:univar_derivation_pxi_x0_g075}
  \end{figure}

  For such a power spectrum with $L \, \sigma_P=150$ and a separation $x=0$, Fig. \ref{fig:univar_derivation_pxi_x0_g075} demonstrates the convergence behaviour of the distribution function. In this case, the calculations with $N=64$ and $N=128$ produce virtually indistinguishable results, so that the former mode number is sufficient for all practical purposes. In general, the steepness of the power spectrum determines which maximum wave number $N$ is necessary: Inserting the wave numbers from Eq. (\ref{eq:pxi_derivation_kn}) into Eq. (\ref{eq:pxi_derivation_gausspwrspec}), we see that $N \sim L \, \sigma_P / 2$ is sufficient for Gaussian power spectra.

 \subsection{Moments}
  \label{sec:univar_moments}
  In this section, we calculate the moments of the distribution. Apart from possible use in future applications, this is also useful as a check for the distribution function derived above, since we can derive the moments in two independent ways and compare results. First, we can get the moments $M_k$ from the derivatives of the characteristic function \citep[p. 63]{Kendall1977}:
  \begin{equation}
   M_k = \iu^{-k} \left. \deriv[k]{\psi(s)}{s} \right|_{s=0} = \iu^{-k} \left. \deriv[k]{}{s} \prod\limits_{n=1}^{\infty} \frac{1}{1 - 2 \iu s C_n} \right|_{s=0} \, .
  \end{equation}
  The first derivative yields the mean of the distribution, or the expectation value of $\xi$:
  \begin{equation}
   \label{eq:moments_mean}
   \overline{\xi} = M_1 = -\iu \left. \deriv{\psi(s)}{s} \right|_{s=0} = 2 \sum\limits_{n=1}^{\infty} C_n \, .
  \end{equation}
  For the variance and other higher-order quantities, we use the central moments, which are the moments of the distribution of $\xi-\overline{\xi}$. The centralised characteristic function is simply
  \begin{equation}
   \psi_c(s) := \left\langle \eto{\iu s \left(\xi-\overline{\xi}\right)} \right\rangle
              = \left( \prod\limits_{n=1}^{\infty} \int \ddiff{g_n}{2} p(g_n) \right) \eto{\iu s \left(\xi-\overline{\xi}\right)}
              = \eto{- \iu s \overline{\xi}} \psi(s) \, ,
  \end{equation}
  and it yields the central moments as \citep[pp. 57, 63]{Kendall1977}
  \begin{equation}
   M_{\mathrm{c}k} = \iu^{-k} \left. \deriv[k]{\psi_c(s)}{s} \right|_{s=0} \,.
  \end{equation}
  The first six non-zero central moments are then
  \begin{equation}
   \label{eq:moments_momc2}
   M_{\mathrm{c}2} = 4 \sum\limits_{n=1}^{\infty} C_n^2 \, .
  \end{equation}
  \begin{equation}
   M_{\mathrm{c}3} = 16 \sum\limits_{n=1}^{\infty} C_n^3 \, .
  \end{equation}
  \begin{equation}
   M_{\mathrm{c}4} =  48  \sum\limits_{n=1}^{\infty} \left(  3C_n^4 +  2C_n^2 \sum\limits_{m    > n}C_m^2 \right) \, .
  \end{equation}
  \begin{equation}
   M_{\mathrm{c}5} = 128  \sum\limits_{n=1}^{\infty} \left( 11C_n^5 +  5C_n^3 \sum\limits_{m \neq n}C_m^2 \right) \, .
  \end{equation}
  \begin{align}
   M_{\mathrm{c}6} = 320  \sum\limits_{n=1}^{\infty} & \left( \ 53C_n^6 + 27C_n^4 \sum\limits_{m \neq n}C_m^2 + 16C_n^3 \sum\limits_{m > n}C_m^3 \right. \nonumber \\
                                                                    & \left. + 18 \sum\limits_{k > m > n}\left(C_nC_mC_k\right)^2 \right) \, . \label{eq:moments_momc6}
  \end{align}  
  From these moments, we can also obtain some conventional statistical quantities: the variance $V(\xi) = M_{\mathrm{c}2}$, the standard deviation $\sigma = \sqrt{V(\xi)}$, the skewness $S(\xi) = M_{\mathrm{c}3}/\sigma^3$ and the kurtosis $K(\xi)~=~M_{\mathrm{c}4}/\sigma^4 - 3$.
  
  Alternatively, we can also calculate the moments from the probability distribution function by the integrals
  \begin{equation}
   M_n = \int\limits_{-\infty}^{\infty} \diff{\xi} \xi^n p(\xi) \, .
  \end{equation}
  An important check for the sanity of the distribution function will be to re-obtain the normalisation as unity by this approach. We can compute it as the moment of order zero:
  \begin{align}
   \mathcal{N} = M_0
   &= \int\limits_{-\infty}^{\infty} \diff{\xi} \sum\limits_{n=1}^{\infty} \mathcal{H}_n \eto{-\xi/(2C_n)} \frac{1}{2C_n} \prod\limits_{m \neq n}^{\infty} \frac{1}{1-\frac{C_m}{C_n}} \nonumber \\
   &= \sum\limits_{n=1}^{\infty} \prod\limits_{m \neq n}^{\infty} \frac{1}{1-\frac{C_m}{C_n}} =: \sum\limits_{n=1}^{\infty} a_n \, . \label{eq:pxi_moments_norm}
  \end{align}
  We can evaluate this sum of products by considering another Fourier transform from $p(\xi)$ back to $\psi(s)$:
  \begin{align}
   \psi(s) &= \int\limits_{-\infty}^{\infty} \diff{\xi} \eto{\iu s \xi} p(\xi) \nonumber \\
           &= \sum\limits_{n=1}^{\infty}  \frac{a_n}{2C_n} \int\limits_{-\infty}^{\infty} \diff{\xi} \mathcal{H}_n \eto{\left[ \iu s - 1/(2C_n) \right] \xi} \nonumber \\ 
           &= \sum\limits_{C_n>0}  \frac{a_n}{2C_n} \int\limits_{0}^{\infty} \diff{\xi} \eto{\left[ \iu s - 1/(2C_n) \right] \xi}  - \sum\limits_{C_n<0}  \frac{a_n}{2C_n} \int\limits_{-\infty}^{0} \diff{\xi} \eto{\left[ \iu s - 1/(2C_n) \right] \xi} \nonumber \\
           &= -\sum\limits_{C_n>0}  \frac{a_n}{2C_n} \frac{1}{\iu s -\frac{1}{2C_n}} - \sum\limits_{C_n<0}  \frac{a_n}{2C_n} \frac{1}{\iu s -\frac{1}{2C_n}} \nonumber \\
           &= -\sum\limits_{n=1}^{\infty} \frac{a_n}{2C_n \left( \iu s -\frac{1}{2C_n} \right)} = \sum\limits_{n=1}^{\infty} \frac{a_n}{1 - 2 \iu s C_n} \, .  \label{eq:pxi_moments_backtrafo}
  \end{align}
  Comparing this expression for the characteristic function to the original from Eq. (\ref{eq:derivation_pxi_charfct}), we get
  \begin{equation}
   \prod\limits_{n=1}^{\infty} \frac{1}{1 - 2 \iu s C_n} = \sum\limits_{n=1}^{\infty} \frac{a_n}{1 - 2 \iu s C_n} \, . \label{eq:pxi_moments_charfcteq}
  \end{equation} 
  Evaluation at $s=0$ yields
  \begin{equation}
   \sum\limits_{n=1}^{\infty} a_n = 1 \,,
  \end{equation}
  so that together with Eq. (\ref{eq:pxi_moments_norm}) we have shown that $\mathcal{N}=1$.
   
  To calculate the higher moments, we make use of the integral
  \begin{equation}
   \int\limits_{0}^{\infty} \diff{\xi} \xi^k H(C) \, \eto{-\xi/(2C)} = 2^{k+1} k! \, C^{k+1} \quad \text{with} \quad k \geq 0
  \end{equation}
  and of sum formulae of the type
  \begin{equation}
   \sum\limits_{n=1}^{N} C_n   \prod\limits_{m \neq n}^N \frac{1}{1-\frac{C_m}{C_n}} = \sum\limits_{n=1}^{N} C_n \, ,
  \end{equation}
  \begin{equation}
   \sum\limits_{n=1}^{N} C_n^2 \prod\limits_{m \neq n}^N \frac{1}{1-\frac{C_m}{C_n}} = \sum\limits_{n=1}^{N} C_n \sum\limits_{m=n}^{N} C_m \, ,
  \end{equation}
  \begin{equation}
   \sum\limits_{n=1}^{N} C_n^3 \prod\limits_{m \neq n}^N \frac{1}{1-\frac{C_m}{C_n}} = \sum\limits_{n=1}^{N} C_n \sum\limits_{m=n}^{N} C_m \sum\limits_{k=m}^{N} C_k \, .
  \end{equation}
  These follow from taking derivatives with respect to $s$ in Eq. (\ref{eq:pxi_moments_charfcteq}) and setting $s=0$. We have checked the results for mean, variance, skewness and kurtosis and have reproduced the results of the characteristic function approach, demonstrating the validity of the probability distribution function.
 
 \vspace{5\lineskip}

 \subsection{Cumulative distribution function}
  \label{sec:univar_cdf}
  From the probability distribution function (\ref{eq:univar_derivation_pxifinal}), we can also directly calculate the cumulative distribution function, defined as $F(\xi) = P( \xi' > \xi)$. For single poles only, it is given by
  \begin{align}
   F(\xi) &= \int\limits_\xi^{\infty} \diff{\xi'} p(\xi') \nonumber \\ 
   &= \sum\limits_{n=1}^{\infty} \frac{1}{2C_n} \prod\limits_{m \neq n} \frac{1}{1-\frac{C_m}{C_n}} \left[ H(\xi) \int\limits_\xi^{\infty} \diff{\xi'} H(C_n) \, \eto{-\xi/(2C_n)} \right. \nonumber \\ 
   &\left. - H(-\xi) \left( \int\limits_\xi^0 \diff{\xi'} + \int\limits_0^{\infty} \diff{\xi'} \right) H(-C_n) \, \eto{-\xi/(2C_n)} \right] \nonumber \\
   &= \sum\limits_{n=1}^{\infty} \left( \mathcal{H}_n \eto{-\xi/(2C_n)} + H(-\xi) \right) \prod\limits_{n \neq m} \frac{1}{1-\frac{C_m}{C_n}} \, . \label{eq:univar_cdf}
  \end{align}
  Again, we use the notation $\mathcal{H}_n = H(\xi)H(C_n) - H(-\xi)H(-C_n)$ for the Heaviside factor, but this time note the extra term of $+ H(-\xi)$. Analogous expressions in the presence of higher-order poles could be obtained by integrating the corresponding probability density (\ref{eq:univar_derivation_pximultipole}).
 
 \vspace{5\lineskip}

 \subsection{A special case - power-law power spectra}
  \label{sec:univar_special}
  In general, the probability distribution function we found is a sum formula that needs to be evaluated numerically. However, if the power spectrum of the underlying random field is a power law $P(k) \propto |k|^{-\nu}$, we can analytically find a more explicit expression for the univariate distribution function. In the case of
  \begin{equation}
   P(k_n) = A |k_n|^{-2} = \frac{L^2A}{4\pi^2n^2} \, ,
  \end{equation}
  with $A$ a normalisation constant, and for a separation of $x=0$, we have $C_n = LA/\left(4\pi^2n^2\right)$ and the product factors are
  \begin{equation}
   a_n = \prod\limits_{m \neq n} \frac{1}{1-\frac{n^2}{m^2}} = \prod\limits_{m \neq n} \frac{1}{\left(1-\frac{n}{m}\right)\left(1+\frac{n}{m}\right)} = 2 (-1)^n \, ,
  \end{equation}
  which is a special case of the infinite product family \citep[p. 754]{Prudnikov1986}
  \begin{equation}
   \label{eq:univar_special_prudnikov}
   \prod\limits_{m=1}^{\infty} \left( 1 - \frac{n^k}{m^k} \right) = - n^{-k} \prod\limits_{m=0}^{k-1} \frac{1}{\Gamma\left(-n\eto{2\pi \iu m / k}\right)} \, .
  \end{equation}
  The probability density function (at zero separation) is then
  \begin{equation}
   p(\xi) = \frac{4\pi^2}{LA} H(\xi) \sum\limits_{n=1}^{\infty} (-1)^{n+1} n^2 \eto{-2\pi^2n^2\xi/(LA)} \, ,
  \end{equation}
  where the field size $L$ comes in through Eq. (\ref{eq:derivation_sigma}). We can now express the cumulative distribution function in terms of known functions as
  \begin{align}
   F(\xi) &= \int\limits_{\xi}^{\infty} \diff{\xi^{\prime}} p(\xi^{\prime})
         = 2 H(\xi) \sum\limits_{n=1}^{\infty} (-1)^{n+1} \eto{-2\pi^2n^2\xi/(LA)} \nonumber \\
        &= H(\xi) \left[ 1 - \vartheta_4\left(0,\eto{-2\pi^2\xi/(LA)}\right) \right] \, .
  \end{align}
  Here, $\vartheta_4\left(0,q\right)$ is a special case of the Jacobi elliptic theta function, which is known to satisfy \citep[p463 ff.]{Whittaker1963}
  \begin{equation}
   \label{eq:univar_special_theta4}
   \sum\limits_{n=1}^{\infty} (-1)^n q^{n^2} = \frac{1}{2} \left[ -1 + \vartheta_4\left(0,q\right) \right] \, .
  \end{equation}
  We can then re-obtain the probability density function by differentiation,
  \begin{align}
   p(\xi) &= -H(\xi) \, \deriv{}{\xi} \vartheta_4\left(0,\eto{-2\pi^2\xi/(LA)}\right) \\ \nonumber
          &= \frac{2\pi^2}{LA} H(\xi) \, \eto{-2\pi^2\xi/(LA)} \vartheta^{\prime}_4\left(0,\eto{-2\pi^2\xi/(LA)}\right) \, ,  \label{eq:univar_special_pxi_dtheta}
  \end{align}
  where $\vartheta^{\prime}_4\left(0,q\right) = \deriv{\vartheta_4\left(0,q\right)}{q}$. It is not clear to us yet whether this connection to elliptical functions is a coincidence for just this special case, or whether it points towards a possible reformulation of the probability distribution for general power spectra.

  However, it allows us to analyse the asymptotic behaviour of the distribution function for large $\xi$. From Eq. (\ref{eq:univar_special_theta4}), we have
  \begin{align}
   \vartheta^{\prime}_4(0,q) &= 2\sum\limits_{n=1}^{\infty} (-1)^n n^2 q^{n^2-1} = - 2 + 8q^3 - 18q^8 + \dots \\ \nonumber
                             &\sim -2 + 8q^3
  \end{align}
  for $q \ll 1$. Then, inserting into Eq. (\ref{eq:univar_special_pxi_dtheta}), we obtain  
  \begin{equation}
   p(\xi) \propto \eto{-2\pi^2\xi/(LA)} \vartheta^{\prime}_4\left(0,\eto{-2\pi^2\xi/(LA)}\right) \propto \eto{-2\pi^2\xi/(LA)}
  \end{equation}
  for $\xi \gg 1$. Thus, the distribution function behaves like a single exponential at large $\xi$, and not like a Gaussian at all. This is in agreement with our general result  $p(\xi) \propto \eto{-\xi / \left(2C_{\mathrm{max}}\right)}$ for large $\xi$ from Sect. \ref{sec:univar_derivation}.

  For the same power spectrum, we can also explicitly calculate the moments. With $C_n \propto P(|k|) \propto |k|^{-2}$, we obtain the $m$-th (central) moment from the sum
  \begin{equation}
   \sum\limits_{n=1}^{\infty} \frac{1}{n^{2m}} = \zeta(2m) = \frac{(-2)^{2m-1}\pi^{2m}}{(2m)!}B_{2m} \, .
  \end{equation}
  Here, $\zeta(m)$ is the Riemann zeta function and $B_m$ are the Bernoulli numbers \citep[p. 776]{Prudnikov1986} given by
  \begin{equation}
   \frac{x}{\eto{x}-1} = \sum\limits_{n=0}^{\infty}B_n\frac{x^n}{n!} \, .
  \end{equation}
  For example, the mean of the distribution then is simply given by $\overline{\xi} = LA/12$ and the standard deviation is $\sigma_{\xi} = LA/\left(6\sqrt{10}\right) \, $.
  
  For power law power spectra with a different exponent, Eq. (\ref{eq:univar_special_prudnikov}) still allows us to express the product factors explicitly in terms of known (trigonometric and hyperbolic) functions. Regrettably, these do not yield any known functions for the full probability distribution, or for the moments, as far as we are aware. Thus, a really explicit form was found for the special case of $P(k) \propto |k|^{-2}$ only.
  
\section{Bivariate distribution}
 \label{sec:bivar}
  
 \subsection{Derivation}
  \label{sec:bivar_derivation}
 
  In this section, we will calculate the bivariate probability distribution function $p(\xi(x_1),\xi(x_2))$, which we will mostly abbreviate as $p(\xi_1,\xi_2)$ with $\xi_1=\xi(x_1)$, $\xi_2=\xi(x_2)$. All preliminaries carry over from the univariate case, and the starting point of this calculation is
  \begin{equation}
            \xi(x_i) = 2 \sum\limits_{n=1}^{\infty} |g_n|^2 \cos ( k_n x_i ) \, .
  \end{equation}
  The characteristic function is now bivariate as well,
  \begin{align}
   \psi(s_1,s_2) &= \left( \prod\limits_{n=1}^{\infty} \int \ddiff{g_n}{2} p(g_n) \right) \eto{\iu \left(s_1\xi_1+s_2\xi_2 \right)} \nonumber \\
                 &= \prod\limits_{n=1}^{\infty} \frac{1}{1 - 2 \iu \left(s_1 C_{n1} + s_2 C_{n2} \right)} \, .
  \end{align}
  In the last step, we have defined a generalised shorthand for the factors $C_{nm} = \sigma_n^2 \cos(k_n x_m)$ to allow for the two different lag parameters $x_m$. It is worth noting at this point that for higher multivariate distributions, say $p(\xi_1,\xi_2,\dots,\xi_k)$, the only change necessary in the characteristic function will be to add additional terms of $s_m C_{nm}$ in this factor, resulting in the generally valid expression
  \begin{equation}
   \label{eq:bivar_derivation_charfctgen}
   \psi(s_1,s_2,\dots,s_k) = \prod\limits_{n=1}^{\infty} \left( 1 - 2 \iu \sum\limits_{m=1}^{k}s_mC_{nm} \right)^{-1} \, .
  \end{equation}
  Next, we will obtain the bivariate probability distribution itself from the characteristic function by Fourier inversion, in analogy to the univariate case. But an important difference arises in this step, since the inversion now contains a double integration. Thus, we have to calculate the following:
  \begin{equation}
   p(\xi_1,\xi_2)=\int\limits_{-\infty}^{\infty} \frac{\fdiff{s_1}}{2\pi} \int\limits_{-\infty}^{\infty} \frac{\fdiff{s_2}}{2\pi} \frac{\eto{-\iu \left(s_1 \xi_1 + s_2 \xi_2 \right)}}{\prod\limits_{n=1}^{\infty} \left[ 1 - 2 \iu \left( s_1 C_{n1} + s_2 C_{n2} \right) \right] } \, .
  \end{equation}
  Since the pairs of variables $(s_1,s_2)$ and $(\xi_1,\xi_2)$ each are mutually independent, the result has to be invariant under exchanging the order of integration. We choose to first integrate over $\diff{s_2}$ and after that over $\diff{s_1}$. From the resulting formula, the symmetry will not be immediately apparent. However, we have checked the equivalence of both approaches by also explicitly evaluating the other choice. Also, we will assume simple poles in both integrations, and will only briefly comment on the effects of multiple poles at the end of this section.

  The poles for the inner integration are now located at
  \begin{equation}
   \label{eq:bivar_derivation_poles}
   s_{2n}=\frac{1}{2 \iu C_{n2}}-\frac{C_{n1}}{C_{n2}}s_1 \, ,
  \end{equation}
  and, for simple poles, their residues are
  \begin{align}
   \operatorname{Res}_{s_{2n}}
     &= \lim\limits_{ s_2 \rightarrow \frac{1}{2 \iu C_{n2}}-\frac{C_{n1}}{C_{n2}}s_1 }
        \left[ \left( s_2-\frac{1}{2 \iu C_{n2}}+\frac{C_{n1}}{C_{n2}}s_1 \right)
        \frac{\eto{-\iu (s_1 \xi_1 + s_2 \xi_2)}}{2\pi} \right. \nonumber \\
     & \left. \times \prod\limits_{m=1}^{\infty} \frac{1}{1 - 2 \iu (s_1 C_{m1} + s_2 C_{m2})} \right] \nonumber \\
     &= \frac{\iu}{2\pi} \frac{\eto{-\xi_2/\left(2C_{n2}\right) - \iu s_1 \left[ \xi_1-\left(C_{n1}/C_{n2}\right)\xi_2 \right]} }{2C_{n2}}
       \left( \prod\limits_{m \neq n} \frac{C_{n2}}{2 \iu D_{nm}} \frac{1}{s_1+\frac{C_{n2} - C_{m2}}{2 \iu D_{nm}}} \right)
  \end{align}
  Here we have simplified the expression by defining the determinant factor
  \begin{equation}
   D_{nm} = C_{n1}C_{m2}-C_{m1}C_{n2} = \det \begin{pmatrix}
                                              C_{n1} & C_{m1} \\
                                              C_{n2} & C_{m2}
                                             \end{pmatrix} \,.
  \end{equation}
  The arguments as to the choice of contours apply exactly as before, since the imaginary part of the poles remains unchanged from Eq. (\ref{eq:derivation_poles}) to Eq. (\ref{eq:bivar_derivation_poles}). Thus, poles with positive $C_{n2}$ factors lie within the lower contour, whereas those with negative $C_{n2}$ lie within the upper contour. The corresponding Heaviside factors encoding this behaviour are $H(\xi_2)H(C_{n2})$ and $H(-\xi_2)H(-C_{n2})$. The winding numbers are $w_n=1$ for the upper and $w_n=-1$ for the lower contour. So we obtain the full integral as
  \begin{align}
    p(\xi_1,\xi_2)
    &= \int\limits_{-\infty}^{\infty} \frac{\fdiff{s_1}}{2\pi} \int\limits_{-\infty}^{\infty} \frac{\fdiff{s_2}}{2\pi} \eto{-\iu (s_1 \xi_1 + s_2 \xi_2)} \prod\limits_{n=1}^{\infty} \frac{1}{1 - 2 \iu (s_1 C_{n1} + s_2 C_{n2})} \nonumber \\
    &= \int\limits_{-\infty}^{\infty} \frac{\fdiff{s_1}}{2\pi} 2 \pi \iu \sum\limits_n w_n \operatorname{Res}_{s_{2n}} \nonumber \\
    &= \sum\limits_{n=1}^{\infty}  \left[ H(\xi_2)H(C_{n2}) - H(-\xi_2)H(-C_{n2}) \right] \eto{-\xi_2/\left(2C_{n2}\right)} \nonumber \\
     & \times \int\limits_{-\infty}^{\infty} \frac{\fdiff{s_1}}{2\pi}
       \frac{\eto{-\iu s_1\left[\xi_1-(C_{n1}/C_{n2})\xi_2\right]}}{2C_{n2}}
       \left( \prod\limits_{m \neq n} \frac{C_{n2}}{2 \iu D_{nm}} \frac{1}{s_1 + \iu \frac{C_{m2} - C_{n2}}{2D_{nm}}} \right) \label{eq:bivar_derivation_firstintegral} \, .
   \end{align}
  \begin{figure*}
   \centering
%    \begin{minipage}[b]{8.5cm}
%      \includegraphics[angle=270,width=1.001\hsize,totalheight=1.001\hsize]{pxixi_x30_y00_g_01_n_16.ps}
%    \end{minipage}
%    \begin{minipage}[b]{8.5cm}
%     \includegraphics[angle=270,width=1.001\hsize,totalheight=1.001\hsize]{pxixi_x40_y50_g_01_n_16.ps}
%    \end{minipage}
%    \includegraphics[width=1.001\hsize]{aa17284-11-fig2.eps}
   \resizebox{0.8\hsize}{!}{\includegraphics{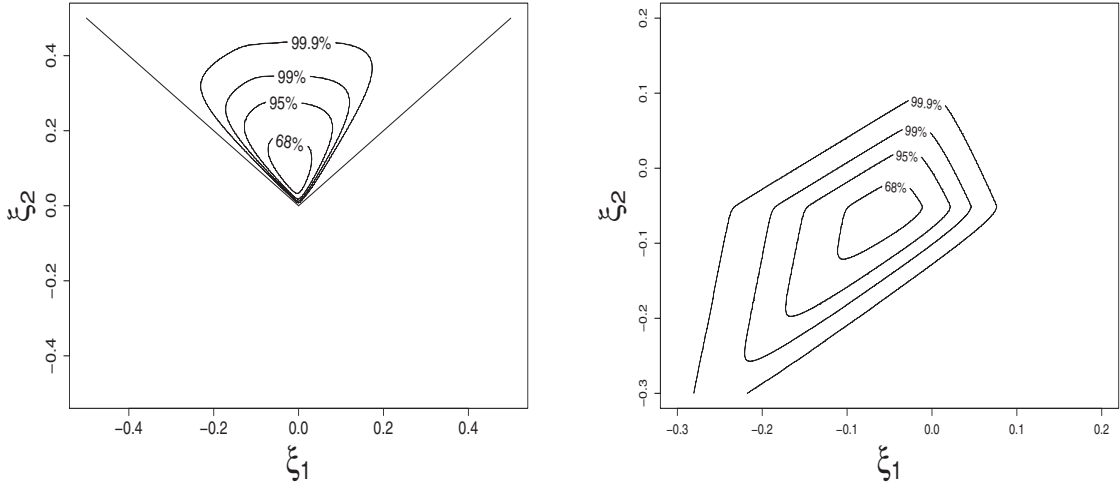}}
   \caption{\small Isoprobability contours of the bivariate distributions $p(\xi_1,\xi_2)$ for a Gaussian power spectrum with $L \, \sigma_P=20$ and 16 modes.
            \newline \textit{Left panel}: $\vec{x}/L=(0.15,0)$, \textit{right panel}: $\vec{x}/L=(0.2,0.25)$.}
   \label{fig:pxixi_gvar01}
  \end{figure*}
   The remaining task is to calculate the second integral. For ease of notation, we will substitute $s_1 \rightarrow s$ and introduce the new variables
   \begin{equation}
    \alpha_n = \xi_1-\frac{C_{n1}}{C_{n2}}\xi_2 \quad \quad \text{and} \quad \quad \beta_{nm} = \frac{C_{m2}-C_{n2}}{2D_{nm}} \, .
   \end{equation}
   Then, the second integral can be reduced to the calculation of the term
   \begin{equation}
    S_n = \int\limits_{-\infty}^{\infty} \frac{\fdiff{s}}{2\pi} \eto{-\iu s \alpha_n}
       \left( \prod\limits_{m \neq n} \frac{1}{s + \iu \, \beta_{nm}} \right) \, .
   \end{equation}
   This integral has poles at $s_{nm} = -\iu \, \beta_{nm}$. For simple poles, the residues are
   \begin{align}
    \operatorname{Res}_{s_{nm}}
     &= \lim\limits_{s \rightarrow - \iu \, \beta_{nm}} \left( s + \iu \, \beta_{nm} \right) \frac{1}{2\pi} \eto{-\iu s \alpha_n} \frac{1}{\prod\limits_{p \neq n} \left( s + \iu \, \beta_{np} \right)} \nonumber \\
     &= \frac{\iu^{-N+2}}{2\pi} \frac{\eto{-\alpha_n \, \beta_{nm}}}{ {\underset{p \neq m}{\underset{p \neq n}{\prod }}} \left( \, \beta_{np}-\beta_{nm} \right)} \, .
   \end{align}
   Furthermore, the choice of contours is also very similar to the previous procedure. We will again close the contour with semi-circles in either the upper or the lower half-plane, and the purely imaginary poles $s_{nm} = -\iu \, \beta_{nm}$ lead, by the same convergence argument, to Heaviside factors $H(\alpha_n)H(\,\beta_{nm})$ for the contour in the lower half-plane and $H(-\alpha_n)H(-\beta_{nm})$ for the contour in the upper half-plane. With the usual winding numbers $w_n = \pm 1$, the integral is
   \begin{align}
    S_n &= 2 \pi \iu \sum\limits_n w_n \operatorname{Res}_{s_{nm}} \nonumber \\
        &= \sum\limits_{m \neq n}  \left[ H(\alpha_n)H(\,\beta_{nm}) - H(-\alpha_n)H(-\beta_{nm}) \right]
           \frac{\iu^{-N+1} \eto{-\alpha_n \, \beta_{nm}}}{\underset{p \neq m}{\underset{p \neq n}{\prod }} \left( \, \beta_{np}-\beta_{nm} \right)} \, .
   \end{align}
   Reinserting this result into the full expression (\ref{eq:bivar_derivation_firstintegral}), the bivariate probability distribution function is
   \begin{align}
    p(\xi_1,\xi_2)_N
     =& \frac{(-1)^{N+1}}{2^N} \sum\limits_{n=1}^{N} \frac{C_{n2}^{N-2}}{ \prod\limits_{m \neq n} D_{nm}} \\ \nonumber
           & \times \sum\limits_{m \neq n} \mathfrak{H}_{nm}
            \frac{\exp\left(\frac{(C_{n2}-C_{m2})\xi_1+(C_{m1}-C_{n1})\xi_2}{2D_{nm}}\right)}
            {\underset{p \neq m}{\underset{p \neq n}{\prod }} \left( \, \beta_{np}-\beta_{nm} \right)} \, ,
   \end{align}
   where we used a shorthand notation for all of the Heaviside factors:
   \begin{align}
    \mathfrak{H}_{nm} =
     & \ \left[ H(\xi_2)H(C_{n2}) - H(-\xi_2)H(-C_{n2}) \right] \nonumber \\
     & \times \left[ H(\alpha_n)H(\,\beta_{nm}) - H(-\alpha_n)H(-\beta_{nm}) \right] \, .
   \end{align}
  Finally, we can bring this expression to a more symmetric form by reinserting the $\beta_{nm}$ and shifting around some factors:
  \begin{align}
   p(\xi_1,\xi_2)_N =&
    \frac{(-1)^{N+1}}{4} \sum\limits_{n=1}^{N} \sum\limits_{m \neq n} \mathfrak{H}_{nm} D_{nm}^{N-3} \\ \nonumber
    & \times \frac{\exp\left(\frac{(C_{n2}-C_{m2})\xi_1+(C_{m1}-C_{n1})\xi_2}{2D_{nm}}\right)}
    { \underset{p \neq m}{\underset{p \neq n}{\prod }} \left( D_{np}-D_{nm}-D_{mp} \right)} \, .
  \end{align}
  This derivation is only valid if, in both integrations, none of the poles vanish or are of higher order. But like we do for the univariate distribution in appendix \ref{sec:multipoles}, we can find corrections to get the most general result. In this case, they become rather unwieldy, and we will not present them in this article, since they can be easily avoided by choosing well-behaved power spectra and non-commensurable lag parameters. However, in Fig. \ref{fig:pxixi_gvar01} we show results for lag parameters which produce such zero modes. The corrections were implemented in the numerical code, as described in Sect. \ref{sec:numerical}, and produce smooth, non-singular results.
  
\begin{figure}
   \resizebox{.95\hsize}{!}{\includegraphics[angle=270]{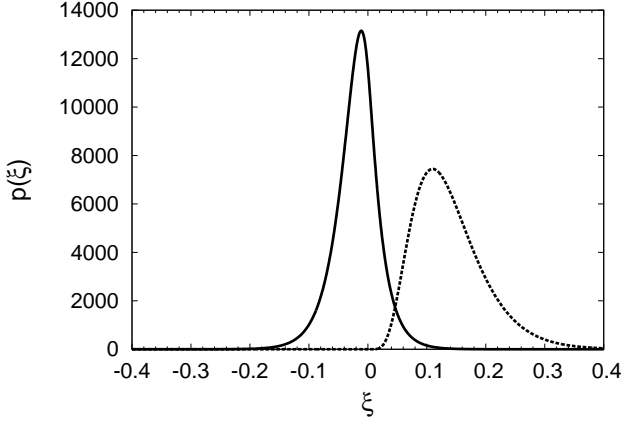}}
   \caption{\small Univariate distributions obtained by marginalisation from the bivariate $p(\xi_1,\xi_2)$, for a Gaussian power spectrum with $L \, \sigma_P=20$, 16 modes and $\vec{x}/L=(0.15,0)$. \textit{solid}: $p(\xi_1)$, \textit{dashed}: $p(\xi_2)$.}
   \label{fig:pxixi_marg}
  \end{figure}

  Both panels show results for an one-dimensional field with a rather narrow Gaussian power spectrum, $L \, \sigma_P=20$, since this allows for convergence with a small number of modes. $N=16$ was used for the diagrams. The left panel shows the distribution for separations $x_1/L=0.15$ and $x_2=0$. It is obvious that the distribution obeys the constraint $|\xi(x_1)| \leq \xi(0)$, with zero probability outside the triangular region defined by this constraint. It is also clearly asymmetric in $\xi_1$, demonstrating the non-Gaussianity. Marginalisations over either $\xi_1$ or $\xi_2$, shown in Fig. \ref{fig:pxixi_marg}, also demonstrate the non-Gaussian nature of the bivariate distribution, and are consistent with our univariate results.

  The right panel of Fig. \ref{fig:pxixi_gvar01} shows the case of $x_1/L=0.2$, $x_2/L=0.25$. Here, no constraints are visible, since we have not fixed any specific value of $\xi_0$, so that the distribution is averaged over all realisations with arbitrary $\xi_0$ and so both $\xi_1$ and $\xi_2$ are unbound. Still, the non-Gaussian nature is apparent in this case as well, since the isoprobability contours have kinks and straight segments following the prescription of the Heaviside terms. 

 \subsection{Moments}
  \label{sec:bivar_moments}
  There are analogues of mean, variance and also of higher moments for multivariate distributions. As we did in Sect. \ref{sec:univar_moments} for the univariate distribution, we can calculate them either by integrating the distribution function or by differentiating the characteristic function. Since two-dimensional integrals are more cumbersome, we restrict ourselves to the characteristic function approach for the bivariate moments.

  The mean is now a vector, but calculation and result are quite similar to the univariate case in Eq. (\ref{eq:moments_mean}):
  \begin{equation}
   \overline{\vec{\xi}} = \begin{pmatrix} \, \overline{\xi_1} \, \\ \, \overline{\xi_2} \, \end{pmatrix} = \left. \begin{pmatrix} -\iu \deriv{\psi(s_1,s_2)}{s_1} \\ -\iu \deriv{\psi(s_1,s_2)}{s_2} \end{pmatrix} \right|_{s_1=0=s_2} = \begin{pmatrix} 2 \sum\limits_{n=1}^{\infty} C_{n1} \\ 2 \sum\limits_{n=1}^{\infty} C_{n2} \end{pmatrix} \, .
  \end{equation}
  For the central moments, the centralised characteristic function is again easy to obtain:
  \begin{equation}
   \psi_c(s_1,s_2) := \left\langle \eto{\iu \vec{s} \left(\vec{\xi}-\overline{\vec{\xi}}\right)} \right\rangle = \eto{- \iu \left(s_1 \overline{\xi_1} + s_2 \overline{\xi_2}\right)} \psi(s_1,s_2) \, .
  \end{equation}
  The analogue of the variance in the bivariate case is the $2 \times 2$ covariance matrix, which we get from the second centralised moments:
  \begin{align}
   \Sigma &= \begin{pmatrix} \Sigma_{11} & \Sigma_{12} \\ \Sigma_{21} & \Sigma_{22} \end{pmatrix}
   = \left. \begin{pmatrix} -\partial_{1}^2 \psi_c(s_1,s_2) & -\partial_{1}\partial_{2} \psi_c(s_1,s_2) \\ -\partial_{2}\partial_{1} \psi_c(s_1,s_2) & -\partial_{2}^2 \psi_c(s_1,s_2) \end{pmatrix} \right|_{s_1=0=s_2} \nonumber \\
   &= \begin{pmatrix} 4 \sum\limits_{n=1}^{\infty} C_{n1}^2 & 4 \sum\limits_{n=1}^{\infty} C_{n1} C_{n2} \\ 4 \sum\limits_{n=1}^{\infty} C_{n1} C_{n2} & 4 \sum\limits_{n=1}^{\infty} C_{n2}^2 \end{pmatrix} \, .
  \end{align}
  Higher central moments $M_{\mathrm{c}k}$ could be obtained by building tensors of rank $k$ from the components
  \begin{equation}
   \iu^{-k} \left. \partial_{j_1} \partial_{j_2} \dots \partial_{j_k} \psi_c(s_1,s_2) \right|_{s_1=0=s_2}  \, .
  \end{equation}

 \subsection{Higher multivariate distributions}
  \label{sec:bivar_multivar}
  In this work, we have derived both univariate and bivariate distribution functions, but no multivariate distributions of more than two correlation functions. In principle, however, these could be obtained by exactly the same type of derivation. From the general multivariate characteristic function, Eq. (\ref{eq:bivar_derivation_charfctgen}), the $k$-variate distribution function can in general be obtained by solving the $k$ integrals in 
  \begin{align}
   p(\xi_1,\xi_2\dots,\xi_k) &= \int\limits_{-\infty}^{\infty} \frac{\fdiff{s_1}}{2\pi} \int\limits_{-\infty}^{\infty} \frac{\fdiff{s_2}}{2\pi} \dots \int\limits_{-\infty}^{\infty} \frac{\fdiff{s_k}}{2\pi} \nonumber \\
   & \times \exp\left(-\iu \sum\limits_{m=1}^{k}s_m\xi_m\right) \prod\limits_{n=1}^{\infty} \left( 1 - 2 \iu \sum\limits_{m=1}^{k}s_mC_{nm} \right)^{-1} \, .
  \end{align}
  Since the integration was already quite complicated for $k=2$, this is not very practical for higher multivariates.

  However, since the multivariate characteristic function is not very complicated and can be computed directly from the input power spectrum, an alternative approach would be to calculate this numerically on a grid in $s$ space, and then do a numerical complex-to-real Fourier transform on these values to obtain $p(\xi)$ on a corresponding grid in $\xi$ space.

  This approach could be quite efficient for practical purposes, where the likelihood function is only needed at discrete points anyway. However, the efficiency and stability of this approach depends strongly on the input $C_{nm}$ factors, since the denominator in the integrand might prove to be numerically hard to handle. Estimating the performance of such a calculation would require further studies.

\section{Numerical implementation}
 \label{sec:numerical}
 The numerical implementation of the probability distribution function, Eq. (\ref{eq:univar_derivation_pxifinal}), or of the cumulative distribution function, Eq. (\ref{eq:univar_cdf}), as required for a Bayesian parameter estimation or even for simply plotting the functions, is not trivial. For a small number of modes and benign parameters (number of dimensions, power spectrum, lag parameter), this can be done straightforwardly in any computer numerics system. In general, however, the summation of many terms with very different orders of magnitude (due to the exponentials and the product factors) leads to a high demand in numerical accuracy. If the internal accuracy of the software is less than the number of significant digits in the summands, rounding and addition errors lead to uncontrollable errors in $p(\xi)$ and all related quantities.
 
 We solved this problem by using the \textsc{arprec} package \citep{Bailey2002}, available for C, C++ and Fortran, which allows calculations with up to 1000 decimal digits. The higher the ratio of separation and field size or the number of significant modes, the higher the necessary precision. However, a large working precision significantly increases the run-time for each evaluation of the likelihood function. Therefore, for a fast and stable implementation suitable for parameter estimation, a way has to be found to determine the necessary precision beforehand.
 
 For plotting the distribution as a function of $\xi$, we only need to compute the product factors once. But in a parameter estimation, the likelihood is evaluated for a different power spectrum in each step, and so an efficient computation of the product factors is also necessary. We will describe one such possibility in Sect. \ref{sec:univar_alternate}, which, however, comes with its own numerical issues. Further investigation into efficient and stable implementations seems necessary.

\section{Discussion}
 \label{sec:discuss}

 \subsection{Analytic properties}
  \label{sec:univar_analytic}
  The probability distribution function, Eq. (\ref{eq:univar_derivation_pxifinal}), has some interesting analytic properties. For lag parameters $x \neq 0$, the distribution (as the full infinite sum) is non-vanishing both for positive and negative $\xi$, and infinitely differentiable at all points. However, there are some subtleties when the sum gets truncated. We can most easily see this for the special case of $x=0$, where all $C_n$ are positive, and we obtain
  \begin{equation}
   \label{eq:univar_analytic_pxi0}
   p(\xi) = H(\xi) \sum\limits_{n=1}^{\infty} \frac{\eto{-\xi/(2C_n)}a_n}{2C_n}  \, ,
  \end{equation}
  with the $a_n$ defined as in Eq. (\ref{eq:pxi_moments_norm}). Thus, we have $p(\xi) > 0$ for $\xi > 0$ and $p(\xi) = 0$ for $\xi \leq 0$. The first derivative is
  \begin{equation}
   p^{\prime}(\xi) = \deriv{p(\xi)}{\xi} = \sum\limits_{n=1}^{\infty} \eto{-\xi/(2C_n)} \frac{a_n}{2C_n} \left( \delta_{\mathrm{D}}(\xi) - \frac{H(\xi)}{2C_n}  \right) \, ,
  \end{equation}
  with the Dirac delta function $\delta_{\mathrm{D}}(x) = \deriv{H(x)}{x}$. At $\xi=0$, we have
  \begin{equation}
   p^{\prime}(0) = \delta_{\mathrm{D}}(0) \sum\limits_{n=1}^{\infty} \frac{a_n}{2C_n} - H(0) \sum\limits_{n=1}^{\infty} \frac{a_n}{4C_n^2} = 0 \, .
  \end{equation}
  For the last step, we had to define $0 \cdot \delta_{\mathrm{D}}(0)=0$, and to use the equality
  \begin{equation}
   \sum\limits_{n=1}^{\infty} \frac{a_n}{(2C_n)^{k+1}} = 0
  \end{equation}
  for $k=0$ and $k=1$, which can be proven by the same argument used for the normalisation in Sect. \ref{sec:univar_moments}, equating two different forms of the characteristic function $\psi(s)$ and then taking the $k$th derivative in $s$.

  So far, we have seen that $p(0)=0$ and $p^{\prime}(0) = 0$. If we consider the full infinite sum for $p(\xi)$, the same is true for all derivatives
  \begin{equation}
   p^{(k)}(\xi) = \deriv[k]{p(\xi)}{\xi} = H(\xi) \sum\limits_{n=1}^{\infty} \eto{-\xi/(2C_n)} \frac{(-1)^k a_n}{(2C_n)^{k+1}} + (\dots) \, ,
  \end{equation}
  where $(\dots)$ stands for terms proportional to $\delta_{\mathrm{D}}(x)$ and its derivatives: We find that $p(\xi)$ can be differentiated infinitely often at all points along the real axis, with all derivatives vanishing at $\xi=0$. But, if the sum is truncated at some arbitrary mode number $N$, this is no longer true. In this case, only the first $N-1$ derivatives will vanish at the origin, and higher derivatives will be discontinuous at this point, making $p(\xi)$ differentiable only $N-1$ times. This also holds for $x \neq 0$: a sum truncated at $N$ modes is only $N-1$ times continuously differentiable. Still, this phenomenon does not harm the convergence of the sum, and the truncated function can still be used as a good approximation of the full sum if $N$ is chosen sufficiently large, as we have demonstrated numerically.

  However, it is of mathematical interest to note that even the infinite sum version of Eq. (\ref{eq:univar_analytic_pxi0}) has peculiar properties at $\xi=0$. If we consider the function $p(\xi)$ on the complex plane, the directional derivatives anywhere but on the real line will not vanish, while they do along the real line. Thus, the function is not analytic in any neighbourhood of $\xi=0$, and therefore the origin is an essential singularity of this function. This phenomenon can also be seen in the special case discussed in Sect. \ref{sec:univar_special}, where 
  \begin{equation}
   p(\xi) = \frac{2\pi^2}{LA} H(\xi) \eto{-2\pi^2\xi/(LA)} \vartheta^{\prime}_4\left(0,\eto{-2\pi^2\xi/(LA)}\right) \, .
  \end{equation}
  The elliptic theta function shares just this analytic behaviour. Still, these functions are smooth for the purposes of real calculus, and there are no problems in practical applications of our results due to this phenomenon.

  Also, discussion in the complex plane allows for another interesting result. Recalling the characteristic function from Eq. (\ref{eq:pxi_moments_backtrafo}), which is in general a complex function, we have
  \begin{equation}
   \Re(\psi(s)) = \sum\limits_{n=1}^{\infty} \frac{a_n}{1 + 4 s^2 C_n^2} \quad \textrm{and} \quad \Im(\psi(s)) = \sum\limits_{n=1}^{\infty} \frac{2 a_n s C_n}{1 + 4 s^2 C_n^2} \, .
  \end{equation}
  Obviously, the real part is an even function in $s$, while the imaginary part is odd. Considering the back transform to the probability distribution function, we find
  \begin{align}
   p(\xi) =& \int\limits_{-\infty}^{\infty} \frac{\fdiff{s}}{2\pi} \eto{-\iu s \xi} \psi(s) \nonumber \\
          =& \int\limits_{-\infty}^{\infty} \frac{\fdiff{s}}{2\pi} \left[\cos(s\xi) - \iu \sin(s\xi)\right] \left[ \Re\left( \psi(s) \right) + \iu \Im\left( \psi(s) \right) \right] \nonumber \\
          =& \int\limits_{-\infty}^{\infty} \frac{\fdiff{s}}{2\pi} \left[\cos(s\xi)\Re\left(\psi(s)\right) + \sin(s\xi)\Im\left(\psi(s)\right)\right] \nonumber \\
           &+ \iu \int\limits_{-\infty}^{\infty} \frac{\fdiff{s}}{2\pi} \left[ \cos(s\xi) \Im\left( \psi(s) \right) - \sin(s\xi) \Re \left( \psi(s) \right) \right] \,.
  \end{align}
  Therefore, the real part of $p(\xi)$ contains an even-even and an odd-odd term, whereas the imaginary part consists of two even-odd terms, which vanish under the integration. This way, we have proven that the probability distribution function is indeed purely real, a fact that was not immediately evident from the original derivation, which had the complex characteristic function as an intermediate step.

  Going to the bivariate distribution function, we find analytical issues similar to those in the univariate case. If both lag parameters are non-zero, $p(\xi_1,\xi_2)$ is non-zero in the full $(\xi_1,\xi_2)$ plane and smooth everywhere. If, however, one of the $\xi_i=0$, then the probability density is strictly zero outside the boundaries defined by the constraint inequality $|\xi(x)| \leq |\xi(0)|$, with the function going smoothly to zero at these boundaries, in the sense that the real directional derivatives (partial derivatives along unit vectors) vanish when crossing the boundaries. However, we have only checked this smoothness numerically, since an analytical calculation of the derivatives would become rather convoluted.

  \begin{figure*}
   \centering
   \begin{minipage}[b]{8.5cm}
    \includegraphics[angle=270,width=1.001\hsize]{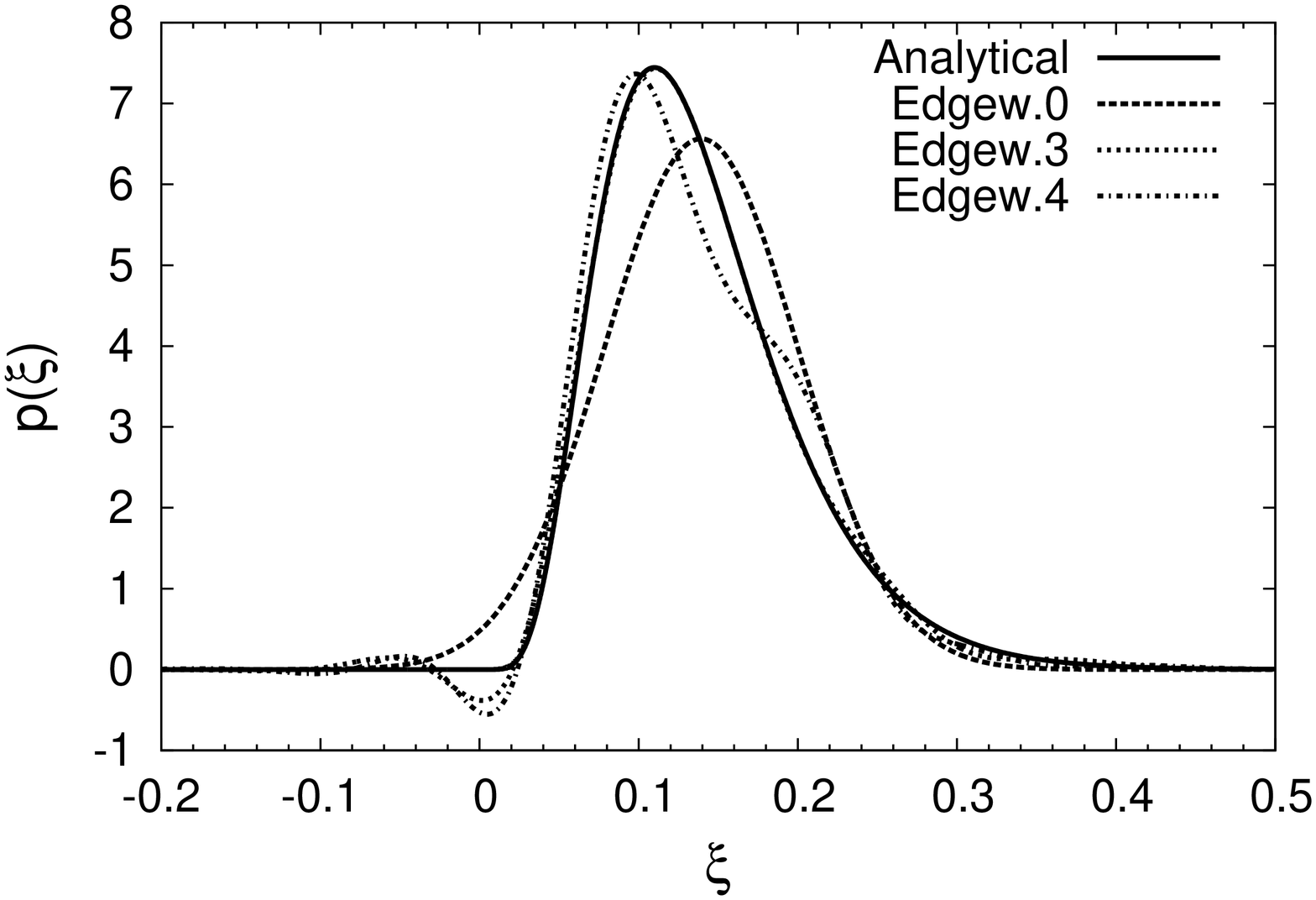}
   \end{minipage}
   \begin{minipage}[b]{8.5cm}
    \includegraphics[angle=270,width=1.001\hsize]{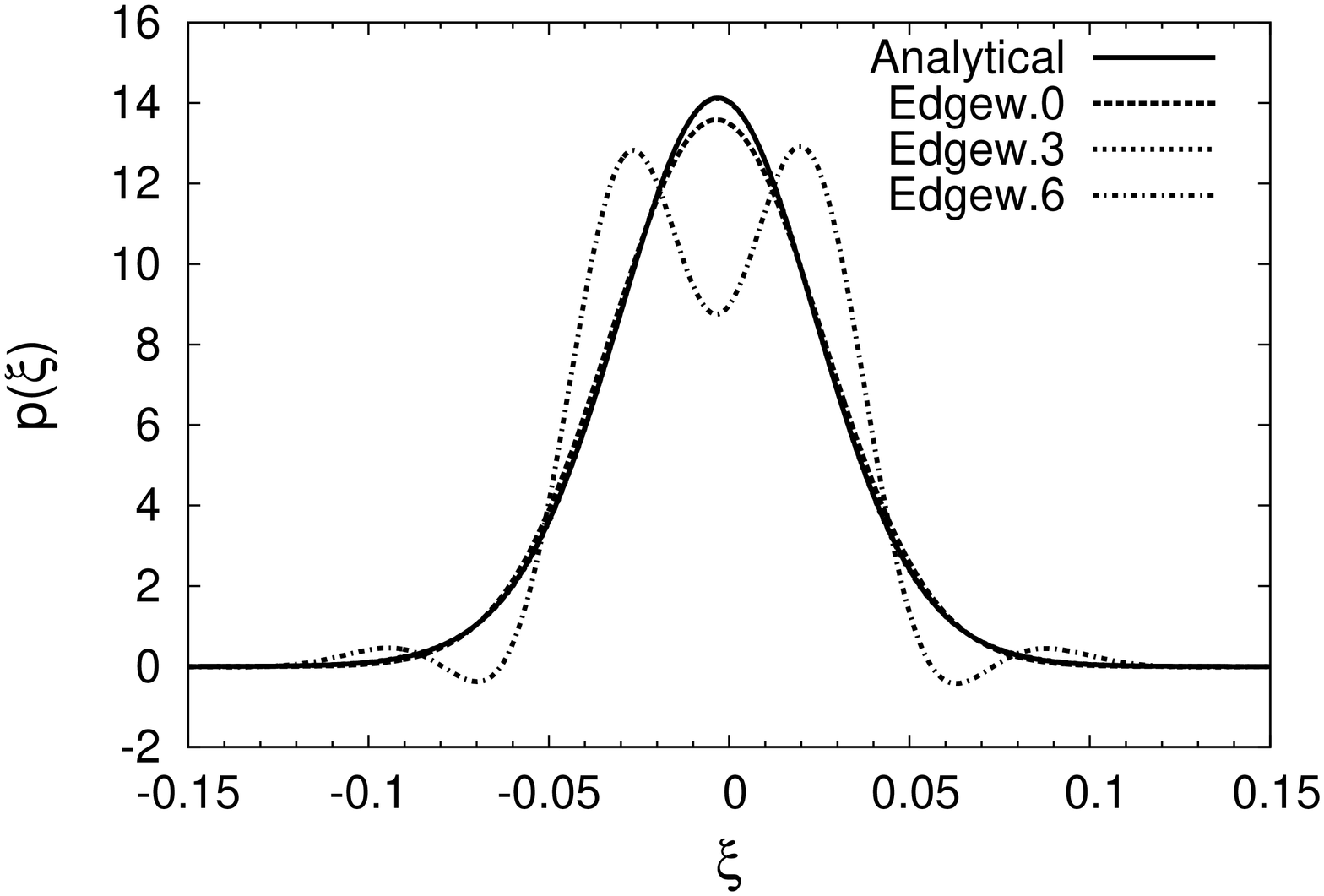}
   \end{minipage}
   \caption{\small Edgeworth expansions compared to full analytical distributions, for Gaussian power spectra.
            \newline \textit{Left panel}: $x=0$, $L \, \sigma_P=20$, $N=16$; \textit{solid}: analytical, \textit{dashed}: 0-order Edg., \textit{dotted}: 3rd order Edg., \textit{dash-dotted}: 4th order Edg.
            \newline \textit{Right panel}: $x/L=0.5$, $L \, \sigma_P=100$, $N=32$; \textit{solid}: analytical, \textit{dashed}: 0-order Edg., \textit{dotted}: 3rd order Edg., \textit{dash-dotted}: 6th order Edg. }
   \label{fig:pxi_edge}
  \end{figure*}
  
 \subsection{Alternative calculation of product factors}
  \label{sec:univar_alternate}
  In principle, the distribution function (\ref{eq:univar_derivation_pxifinal}) is simply a weighted sum of exponentials. However, the weights are given by the products
  \begin{equation}
   a_n = \prod\limits_{m \neq n} \frac{1}{1-\frac{C_m}{C_n}} \, .
  \end{equation}
  These make the summation complicated, since closed-form analytic expressions exist in special cases only, and high accuracy is needed to compute them numerically. Therefore, we were looking for alternative ways to calculate these products, and found an approach by a linear set of equations (LSE).

  We obtain the first equation of this system from the normalisation, as calculated in Sect. \ref{sec:univar_moments}:
  \begin{equation}
   \int\limits_{0}^{\infty} \diff{\xi} p(\xi) = \sum\limits_{n=1}^N a_n = 1 \, .
  \end{equation}
  For the remaining equations, we consider again the smoothness of the distribution function at the origin. As already discussed in Sect. \ref{sec:univar_analytic}, the first $N-1$ derivatives of $p(\xi)$ with respect to $\xi$ yield the equations
  \begin{equation}
   \label{eq:prodfac_derivatives}
   p^{(k)}(0) = \sum\limits_{n=1}^N \frac{a_n}{(2C_n)^{k+1}} = 0 \, .
  \end{equation} 
  Then, we can consider the $N$ factors $a_n$ as the components of a vector $\vec{a} = (a_1,a_2,\dots,a_N)$. Combining this with another vector $\vec{b}=(1,0,0,\dots,0)$ with $N$ entries and with the $N \times N$ coefficient matrix $\tens{M}$ of the above equations, we obtain the LSE
  \begin{equation}
   \begin{pmatrix}
    1           & 1           & \dots  & 1 \\
    (2C_1)^{-1} & (2C_2)^{-1} & \dots  & (2C_N)^{-1} \\
    (2C_1)^{-2} & (2C_2)^{-2} & \dots  & (2C_N)^{-2} \\
    \vdots      & \vdots      & \ddots & \vdots \\
    (2C_1)^{-N} & (2C_2)^{-N} & \dots  & (2C_N)^{-N} \\
   \end{pmatrix}
   \cdot \begin{pmatrix}
          a_1 \\
          a_2 \\
          a_3 \\
          \vdots \\
          a_N
   \end{pmatrix}
   = \begin{pmatrix}
      1 \\
      0 \\
      0 \\
      \vdots \\
      0
     \end{pmatrix} \, .
  \end{equation}
  We can then solve this LSE by any of the usual methods. In particular, we applied the Singular Value Decomposition (SVD) as implemented in the GNU Scientific Library \citep[GSL, ][]{Galassi2009}. This implementation has limited precision and therefore runs into the same numerical problems described in Sect. \ref{sec:numerical}. However, SVD solutions for many numerically challenging problems exist, and therefore the general approach seems promising. 

 \subsection{Edgeworth expansion}
  \label{sec:univar_edge}
  Since the moments have simpler expressions (Eqs. \ref{eq:moments_mean} and \ref{eq:moments_momc2} to \ref{eq:moments_momc6}) than the probability distribution function itself, it seems promising to express the distribution in terms of its moments. One way to do so is the Edgeworth asymptotic expansion, described in \cite{Blinnikov1998}. It has the form
  \begin{align}
   p\left(\sigma\xi+\overline{\xi}\right) = \frac{\eto{-\xi^2/2}}{\sqrt{2\pi}\sigma} & \left[ 1 + \sum\limits_{n=1}^{\infty} \sigma^n \sum\limits_{\{k_m\}} \frac{\operatorname{He}_{n+2r}(\xi)}{k_m!} \right. \label{eq:edgeworth} \\
         & \left. \times \prod\limits_{m=1}^n \left( \frac{\kappa_{m+2}}{\sigma^{2m+2}(m+2)!} \right)^{k_m} \right] \nonumber
  \end{align}
  with the (probabilist's) Hermite polynomials $\operatorname{He}_n$, and $\overline{\xi}$ and $\sigma$ as given in Sect. \ref{sec:univar_moments}. The inner sum runs over all sets $\{k_m\}$ of non-negative integers solving the Diophantine equation
  \begin{equation}
   \sum\limits_{m=1}^{n}mk_m=n \, ,
  \end{equation}
  and we also defined, for each such set,
  \begin{equation}
   r = \sum\limits_{m=1}^{n}k_m \, .
  \end{equation}
  We calculate the $\{k_m\}$ and $r$ with the algorithm presented in Appendix C of \cite{Blinnikov1998}.

  The first term in the Edgeworth expansion is a simple Gaussian, and the higher order terms are given by the cumulants $\kappa_n$ of the distribution in question, which we can derive from the central moments $M_{\mathrm{c},n}$ as
  \begin{equation}
   \kappa_n = M_{\mathrm{c},n} - \sum\limits_{m=1}^{n-1} \binom{n-1}{m-1} \kappa_m M_{\mathrm{c},n-m} \, .
  \end{equation}
  In Fig. \ref{fig:pxi_edge}, we demonstrate the performance of the Edgeworth expansion in two examples, both for Gaussian power spectra. The left panel shows a distribution at zero lag, with $L \, \sigma_P=20$ and 16 modes. The Edgeworth term of order zero, a Gaussian with the same mean and variance as the full distribution, is a very bad fit in this case. The third-order Edgeworth expansion is the best fit, fitting the peak of the distribution almost perfectly and the tail decently well, while also producing negative probabilities for some $\xi<0$. Adding additional terms only makes the approximation worse, distorting it in the high probability region. For the right panel, we used $x/L=0.5$, $L \, \sigma_P=100$ and $N=32$. For this more symmetric distribution, the Gaussian is already a better fit; still, the third order Edgeworth expansion fits even better at the peak. Higher orders again begin to deviate, as evidenced by the strongly two-peaked sixth-order expansion.

  \begin{figure*}
   \centering
   \begin{minipage}[b]{8.5cm}
    \includegraphics[angle=270,width=1.001\hsize]{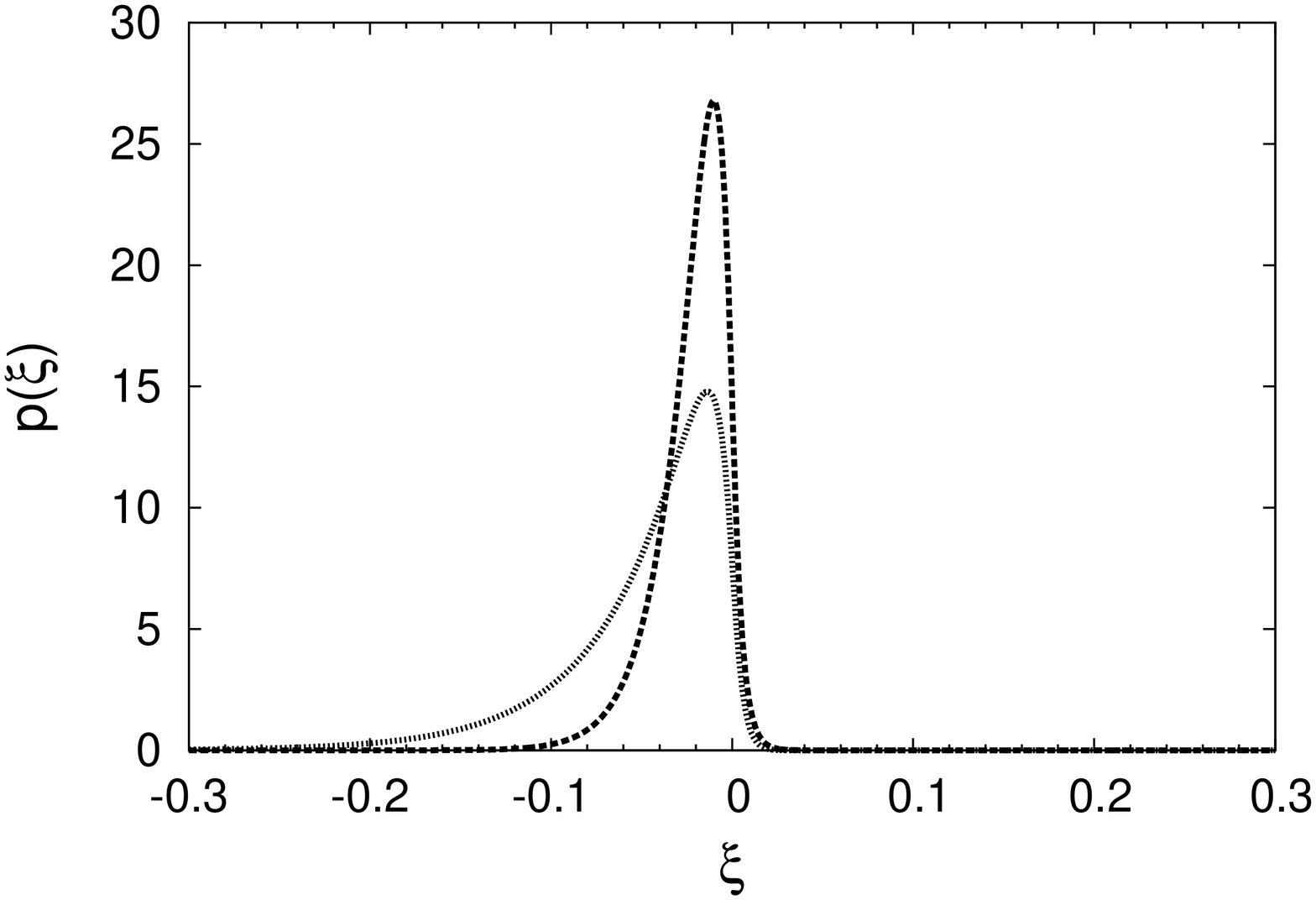}
   \end{minipage}
   \begin{minipage}[b]{8.5cm}
    \includegraphics[angle=270,width=1.001\hsize]{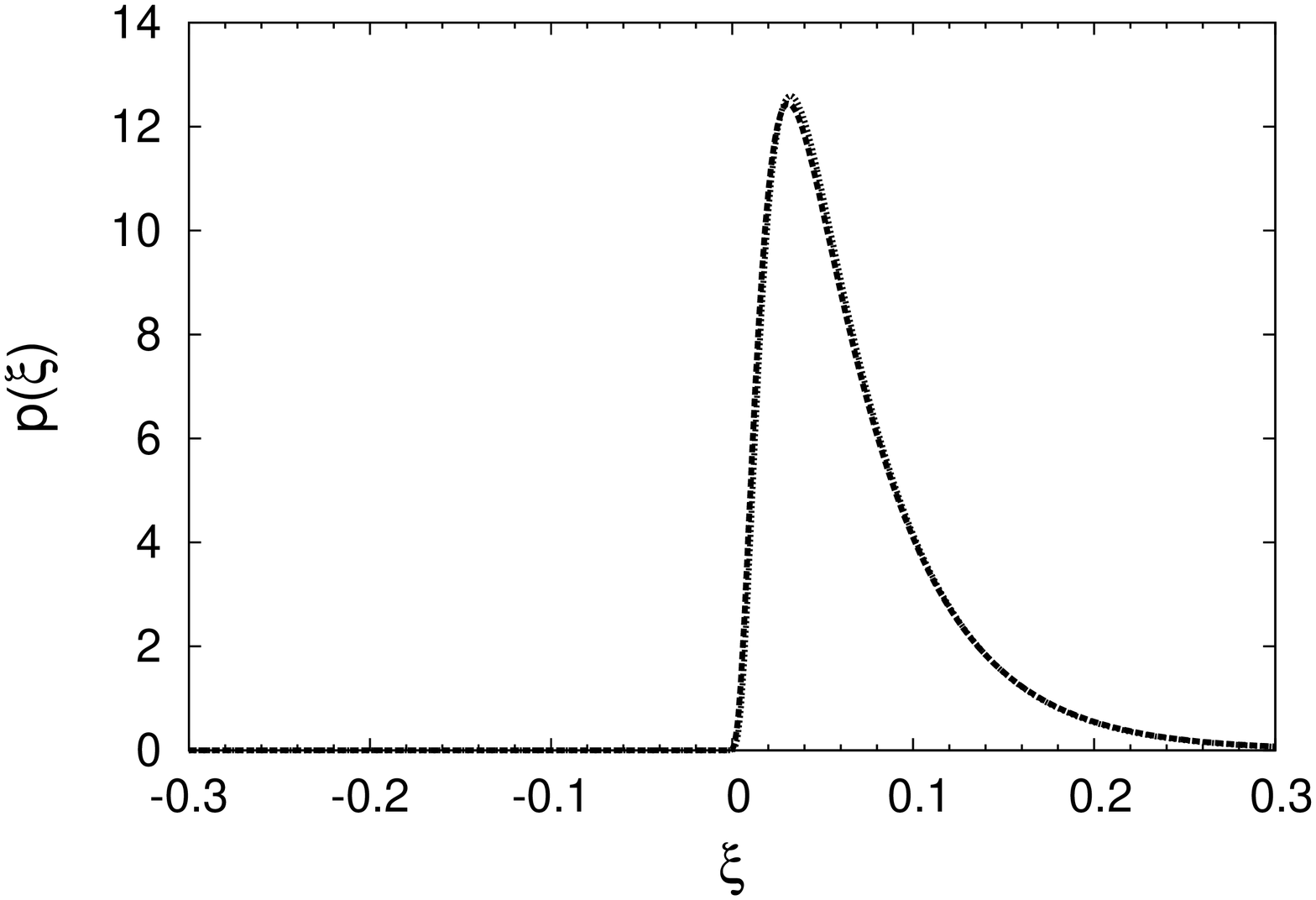}
   \end{minipage}
   \caption{\small Univariate distributions for a power-law power spectrum and different separation vectors $\vec{x}$, demonstrating isotropy in 2D.
    \newline \textit{Left panel}: \textit{dashed}: $\vec{x}/L=(0.3,0)$, \textit{dotted}: $\vec{x}^\prime/L\approx(0.212,0.212)$, \textit{Right panel}: \textit{dashed}: $\vec{x}/L=(0.03,0)$, \textit{dotted}: $\vec{x}^\prime/L\approx(0.0212,0.0212)$}
   \label{fig:pxi_2dim_isotropy}
  \end{figure*}
  
  The reason for this behaviour is that, for the higher terms, the Edgeworth expansion assigns high weights to the tails of the distribution. Since the distribution we are analysing is in fact far from Gaussian, especially in the tails, the fit to the peak correspondingly gets worse, and often even multiple peaks appear. This is similar to the effect noticed by \cite{Blinnikov1998} for the example of $\chi^2$ distributions. Also, the regions of negative probability density are generic features of the Edgeworth expansion when applied to strongly non-Gaussian distributions.
  
  Still, if we truncate the Edgeworth expansion after a wisely chosen number of terms, it will provide a good approximation to the true probability distribution. Since Eq. (\ref{eq:edgeworth}) is an asymptotic series expansion, it generally does not converge, but we do have a method at hand to find the optimal number of terms and to control the error we make. If we truncate the sum over $n$ in Eq. (\ref{eq:edgeworth}) after $N$ terms, the last term retained will also give the order of the difference between the full $p(\xi)$ and this partial sum expansion. Therefore we can, in practice, simply truncate the expansion at the term with minimal contribution (measured at the peak of the distribution, at a certain point of evaluation, or integrated over a domain in $\xi$ we are interested in).

  For both cases illustrated in Fig. \ref{fig:pxi_edge}, this criterion gives $N=3$ as the optimal order of expansion. Also for other parameters $x/L$ and $L \, \sigma_P$, such a third-order Edgeworth expansion seems to be a safe choice to get a substantial improvement as opposed to a simple Gaussian likelihood.

 \subsection{Higher dimensions} 
  \label{sec:univar_ndim}
  In all of the above calculations, we assumed one-dimensional random fields. However, we can easily generalise all results to higher dimensions. If we go to $N_{\mathrm{dim}}$ dimensions, with lag parameters $\vec{x} = (x_1, \dots, x_{N_{\mathrm{dim}}})$, then the allowed Fourier modes are all
  \begin{equation}
   \vec{k} = \frac{2\pi}{L} \vec{n} = \frac{2\pi}{L} (n_1, \dots, n_{N_{\mathrm{dim}}})
  \end{equation}
  with integer $n_i$. Still, all the modes are independent and each has a Gaussian probability distribution with its dispersion given by $\sigma_{\vec{n}}=P(|\vec{k}_{\vec{n}}|)/L^N_{\mathrm{dim}}$. The derivation of $p(\xi)$ stays exactly the same as presented in Sect. \ref{sec:univar_derivation}. Where necessary, we can renumber all modes $\vec{n}$ with a single integer $n$ by an arbitrary scheme and retain the old notations with scalar indices. Then, Eq. (\ref{eq:univar_derivation_pxifinal}) still holds, with only two important changes.

  First, the sums now go over many more modes, namely all possible vectors of integers $\vec{n} = (n_1, \dots, n_{N_{\mathrm{dim}}})$ with $n_i \in \mathbb{N}$. If, in a numerical implementation, we want to cover a box in $k$-space with $N$ grid points in each direction, we therefore end up with $N^{N_{\mathrm{dim}}}$ modes. Besides the increased computational cost, this also leads to frequent occurrences of multiple poles. This is the case especially for $\vec{x} = \vec{0}$, since then the $C_{\vec{n}}$ depend on $|\vec{n}|$ only. Therefore, the generalised result of appendix \ref{sec:multipoles} gets naturally important in higher dimensions. However, in real application scenarios, where accuracy is limited by external factors anyway, it is always possible to avoid multiple poles by slightly changing the $C_{\vec{n}}$ factors, e.g. by adding a small number $\epsilon$ in the cosine,
   \begin{equation}
    C_{\vec{n}} \rightarrow C_{\vec{n}}^\prime = \sigma_{\vec{n}}^2 \cos(\vec{x} \cdot \vec{k}_{\vec{n}} + \epsilon) \, .
   \end{equation}
  Doing this removes the multiple poles while only slightly changing the results with respect to using the unmodified $C_n$ and the full multi-pole formula, Eq. (\ref{eq:univar_derivation_pximultipole}).

  As a second effect, the factors $C_n$ now depend on the angle between separation vector $\vec{x}$ and mode vector $\vec{k}_{\vec{n}}$:
  \begin{equation}
   C_{\vec{n}} = \sigma_{\vec{n}}^2 \cos(\vec{x} \cdot \vec{k}_{\vec{n}}) \, .
  \end{equation}
  However, a Gaussian random field is completely determined by its power spectrum, and when we assume a $P(\vec{k}_{\vec{n}})$ which depends on the absolute values of the $\vec{k}_{\vec{n}}$ only, such a field should be statistically isotropic. Any anisotropies seen in our results must be a consequence of using a finite, cubic field instead of an infinite field. So we expect that all anisotropies vanish as soon as most of the power comes from scales much smaller than the field size. For power spectra concentrated narrowly towards $\vec{k}=\vec{0}$, i.e. those with only a few significant modes, the low-$k$, large-scale modes will dominate for any finite field size. But for a wide power spectrum, where many small-scale modes contribute, a sufficiently large field size should lead to approximate isotropy.
   
  This is demonstrated for a two-dimensional field with a power spectrum $P(k) = k^{-2}$ in Fig. \ref{fig:pxi_2dim_isotropy}. Both panels show the probability distributions for two separation vectors $\vec{x}$ and $\vec{x}^\prime$ of the same length, but rotated by $45^{\circ}$. (Or, more precisely, with $\vec{x}=(x_1,0)$ and $\vec{x}^\prime=\left(x_1\cos(45^{\circ}+\epsilon), x_1\cos(45^{\circ}-\epsilon)\right)$, with $\epsilon=10^{-10}$, to avoid double poles.) For a large separation to field size ratio, $|\vec{x}|/L=0.3$, seen in the left panel, the distributions for different separation vectors are quite different, while for $|\vec{x}|/L=0.03$, in the right panel, they have almost converged. Note that, for $P(k) = k^{-2}$ in 2 dimensions, the field size $L$ cancels out of the $\sigma_n$, and therefore we could keep the power spectrum normalisation constant while changing $|\vec{x}|/L$, without changing the scale of $p(\xi)$.

  Isotropy is even easier to obtain for power spectra that are small both for very small and very large $k$ and large only at intermediate wavelengths, since then the low-$k$, large scale modes do not harm isotropy. The $\Lambda$CDM power spectrum in cosmology fulfils this condition, allowing isotropic N-body simulations with reasonable field sizes.
                                                                                                                                                                                                                                                                                                                                                                                                                                                                     
  However, these are no fundamental changes, and the resulting distribution function has all the same properties as the one-dimensional version. So all our results can be readily applied to higher dimensions, as long as the computational difficulties can be handled.

\section{Conclusions}
 \label{sec:conclusions}
 We have considered the problem of accurate likelihood functions for Bayesian analyses of data from Gaussian random fields. Making use of Fourier mode expansions and characteristic functions, we have derived analytically the probability distribution function of the correlation function for a one-dimensional finite Gaussian random field. For general power spectra, we can only give a sum formula for the distribution function. However, for the special power spectrum $P(k)\propto k^{-2}$, we have found an explicit expression in terms of elliptic theta functions. We can also, for general power spectra, calculate arbitrary moments and cumulants of the distribution by much simpler sum expressions.

 Then, we continued the analytical approach for bivariate distributions as well, and found a similar, but even more complicated sum formula as a result. We have also outlined a general procedure for calculating arbitrary high multivariate distributions, though this is not feasible in practice.

 Furthermore, we have considered the analytical properties of the new probability distribution function, which are well understood and consistent with numerical results. We used the moments of the univariate distribution to construct an Edgeworth expansion, which is able to closely approximate the true distribution function in the region of highest likelihood, and therefore could be a useful replacement of simple Gaussian approximations in Bayesian analyses. Finally, we found that all our results easily generalise to multi-dimensional fields.

 Considering possible future applications of these results, we have to point out the importance of improving the correlation function likelihood, as well as the further steps that are necessary for a practical implementation.

 As already mentioned in the introduction, the Gaussian approximation for the likelihood can lead to considerable deviations in parameter estimation. For example, in cosmic shear studies, this leads to significantly reduced accuracy \citep{Hartlap2009}. Similar effects are to be expected in other fields where correlation functions are used.

 However, the work presented in this article has so far been purely mathematical, and the results are not readily applicable to real data. The main obstacle lies in the infeasibility of analytical calculations for higher multivariate distributions. If data of the correlation function over $N$ bins needs to be analysed, we would need the full $N$-variate distribution function. Therefore, we expect that a numerical approach, as by \cite{Wilking2011}, is best suited for practical computations. Still, their 'quasi-Gaussian' approach makes direct use of the analytical univariate distribution function presented in this article. We also expect that the analytical results will yield important guidance and cross-checks for future numerical implementations.

 We also note that our analytical results depend on the assumption of a Gaussian random field, whereas a lot of cosmological data probes the evolved density field on small scales, which is far from Gaussian. Nevertheless, applications for the analytical likelihood function could be found on very large scales, or in cosmic microwave background analysis, since the density fluctuations at that epoch were still either Gaussian or close to Gaussian. Furthermore, this work could also be relevant to fields outside of cosmology, for example in the common problem of time series analysis of Gaussian random processes.

\begin{acknowledgements}
 We thank Jan Hartlap, Cristiano Porciani and Philipp Wilking for their help, comments and discussion during this project. This work was supported by the Deutsche Forschungsgemeinschaft under the project SCHN~342/11--1. D. Keitel was also supported by the Bonn-Cologne Graduate School of Physics and Astronomy. 
\end{acknowledgements}

\appendix
\section{Multiple poles}
 \label{sec:multipoles}
 So far, to keep calculations simple, we have considered simple poles only. However, it is entirely possible to have multiple poles in the characteristic function, i.e. to have some mode numbers $\vec{m} \neq \vec{n}$ with $C_{\vec{m}} = C_{\vec{n}}$. This could happen if
 \begin{itemize}
  \item the power spectrum is non-monotonic, with $P(\vec{k}_{\vec{n}}) = P(\vec{k}_{\vec{m}})$ for some $m \neq n$,
  \item in higher dimensions, several different modes $\vec{n}$ have identical absolute value $|\vec{k}_{\vec{n}}|$,
  \item one of the lag parameters is commensurable with $\frac{\pi}{2}$ and thus produces periodicity in $\cos(\vec{x}\cdot\vec{k}_{\vec{n}})$.
 \end{itemize}
 From now on, we will use a scalar index $n$ running over all modes $\vec{n}$ by some arbitrary numbering scheme. If a multiple pole of order $k$ is present for some set of modes, which we will call $\mathbb{N}^*_n = \{ m \in \mathbb{N} | C_m = C_n \}$, we can calculate the residue as
  \begin{equation}
   \operatorname{Res}_{s_n} = \tfrac{1}{N_n(k-1)!} \lim\limits_{s \rightarrow s_n} \pderiv[k-1]{}{s} \left( \frac{(s-s_n)^k \eto{-\iu s \xi}}{\left(1-2 \iu s C_n\right)^k} \prod\limits_{C_m \neq C_n} \frac{1}{1-2 \iu s C_m} \right) \, ,
  \end{equation}
 where the weight $N_n$ is the number of modes in $\mathbb{N}^*_n$. We can then rewrite $p(\xi)$ as the usual sum, limited to single pole modes, plus correction terms for the multiple poles. For example, when there are some double poles $\mathbb{N}^*$ and all other modes have single poles, the full expression reads
 \begin{align}
  p(\xi) = & \  \sum\limits_{n \in \mathbb{N} \slash \mathbb{N}^*}  \frac{\mathcal{H}_n \eto{-\xi/(2C_n)}}{2C_n} \prod\limits_{m \neq n} \frac{1}{1-\frac{C_m}{C_n}} \nonumber \\
           & + \sum\limits_{n \in \mathbb{N}^*}  \frac{\mathcal{H}_n \eto{-\xi/(2C_n)}}{4C_n^2} \prod\limits_{m \neq n} \frac{1}{1-\frac{C_m}{C_n}}
            \left( \xi - 2 \sum\limits_{C_m \neq C_n} \frac{C_m}{1-\frac{C_m}{C_n}} \right)  \, .
 \end{align} 
 For general types of poles, we can absorb all multi-pole contributions in a pole order correction factor $\mathcal{P}_n$, so that we can write the probability distribution function compactly as
 \begin{equation}
  \label{eq:univar_derivation_pximultipole}
  p(\xi) =   \sum\limits_{n=1}^{\infty}  \mathcal{H}_n \eto{-\xi/(2C_n)} \, \mathcal{P}_n \prod\limits_{m \neq n} \frac{1}{1-\frac{C_m}{C_n}}  \, .
 \end{equation}    
 If the $n$-th mode belongs to a set of poles of order $k$, its correction factor is
 \begin{equation}
  \label{eq:univar_derivation_pxinpole}
  \mathcal{P}(k) = \frac{1}{k!(2C_n)^k} \left( \sum\limits_{ \left\{ a_i \right\} } A_{\left\{ a_i \right\}} \xi^{a_0} \prod\limits_{i=1}^{k-1} \left[ \sum\limits_{C_m \neq C_n} \left(\frac{C_m}{1-\frac{C_m}{C_n}}\right)^i \right]^{a_i} \right) \, ,
 \end{equation}
 where the outer sum goes over all sets of integers $a_i$ that fulfil
 \begin{equation}
  a_0 + \sum\limits_{i=1}^{k-1} a_i i = k-1 \, .
 \end{equation}
 This expression was extrapolated from explicit calculation of up to quintuple poles. The prefactors $A_{\left\{ a_i \right\}}$ obtained in these calculations are given in table \ref{tbl:pole_order_corrfct}. We still have to find a general expression for these prefactors, so that we can also give $\mathcal{P}(k)$  for $k>5$.
 
 \begin{table}
  \caption{\small Pole order correction factors}
  \label{tbl:pole_order_corrfct}
  \centering
  \begin{tabular}{r r r}
   \hline\hline
   $k$ & $\left\{ a_i \right\}$ & $A_{\left\{ a_i \right\}}$ \\
   \hline
   1 & 0, 0, 0, 0, 0 &    1 \\
   \hline
   2 & 1, 0, 0, 0, 0 &    1 \\
     & 0, 1, 0, 0, 0 & -  2 \\
   \hline
   3 & 2, 0, 0, 0, 0 & -  1 \\
     & 1, 1, 0, 0, 0 &    4 \\
     & 0, 2, 0, 0, 0 & -  4 \\
     & 0, 0, 1, 0, 0 & -  4 \\
   \hline
   4 & 3, 0, 0, 0, 0 & -  1 \\
     & 2, 1, 0, 0, 0 &    6 \\
     & 1, 2, 0, 0, 0 & - 12 \\
     & 1, 0, 1, 0, 0 & - 12 \\
     & 0, 3, 0, 0, 0 &    8 \\
     & 0, 1, 1, 0, 0 &   24 \\
     & 0, 0, 0, 1, 0 &   16 \\
   \hline
   5 & 4, 0, 0, 0, 0 &    1 \\
     & 3, 1, 0, 0, 0 & -  8 \\
     & 2, 2, 0, 0, 0 &   24 \\
     & 2, 0, 1, 0, 0 &   24 \\
     & 1, 3, 0, 0, 0 & - 32 \\
     & 1, 1, 1, 0, 0 & - 64 \\
     & 0, 4, 0, 0, 0 &   16 \\
     & 0, 2, 1, 0, 0 &   96 \\
     & 0, 1, 0, 1, 0 &  128 \\
     & 0, 0, 2, 0, 0 &   48 \\
     & 0, 0, 0, 0, 1 &   96 \\
   \hline
  \end{tabular}
 \end{table}

\section*{Note added in proof}
 Recently, it was brought to our attention (thanks to Stefan Hilbert) that a related distribution has long been known in the fields of renewal theory and signal processing. In the special case that all $C_n > 0$, the correlation function (Eq. \ref{eq:univar_derivation_xi}) simplifies to a sum of squares of absolute values of complex Gaussian random variables, equivalent to a sum of exponential variables, and our univariate distribution (Eq. \ref{eq:univar_derivation_pxifinal}) is equivalent to a type of generalized Erlangian distribution \citep[see][]{Cox1962}. A more generalized, multivariate version is given as 'Theorem 4' in \cite{Hammarwall2008}, but this is still limited to positive parameters. Therefore, these results do not apply for the case of arbitrary signs of the $C_n$ which is needed for our purpose.

\bibliographystyle{aa} % style aa.bst
\bibliography{aa17284-11_bib} % your references Yourfile.bib

\begin{thebibliography}{15}
\expandafter\ifx\csname natexlab\endcsname\relax\def\natexlab#1{#1}\fi

\bibitem[{Bailey {et~al.}(2002)Bailey, Hida, Li, \& Thompson}]{Bailey2002}
Bailey, D.~H., Hida, Y., Li, X.~S., \& Thompson, O. 2002, ARPREC: An arbitrary
  precision computation package, Tech. rep.

\bibitem[{{Blinnikov} \& {Moessner}(1998)}]{Blinnikov1998}
{Blinnikov}, S. \& {Moessner}, R. 1998, \aaps, 130, 193

\bibitem[{Cox(1962)}]{Cox1962}
Cox, D. 1962, Renewal theory, Methuen's monographs on applied probability and
  statistics (Methuen, London)

\bibitem[{Fu {et~al.}(2008)}]{Fu2007}
Fu, L. {et~al.} 2008, Astron. Astrophys., 479, 9

\bibitem[{Galassi {et~al.}(2009)Galassi, Davies, Theiler, Gough, Jungman,
  Alken, Booth, \& Rossi}]{Galassi2009}
Galassi, M., Davies, J., Theiler, J., {et~al.} 2009, {GNU scientific library:
  reference manual}, 3rd edn. (Network Theory, Bristol)

\bibitem[{Hammarwall {et~al.}(2008)Hammarwall, Bengtsson, \&
  Ottersten}]{Hammarwall2008}
Hammarwall, D., Bengtsson, M., \& Ottersten, B. 2008, IEEE Transact. Signal
  Proc., 56, 1188

\bibitem[{{Hartlap} {et~al.}(2009){Hartlap}, {Schrabback}, {Simon}, \&
  {Schneider}}]{Hartlap2009}
{Hartlap}, J., {Schrabback}, T., {Simon}, P., \& {Schneider}, P. 2009, \aap,
  504, 689

\bibitem[{Kendall \& Stuart(1977)}]{Kendall1977}
Kendall, M. \& Stuart, A. 1977, {The advanced theory of statistics. Vol. 1:
  Distribution theory}, 4th edn. (Charles Griffith \& Company, London)

\bibitem[{Okumura {et~al.}(2008)}]{Okumura2007}
Okumura, T. {et~al.} 2008, Astrophys. J., 676, 889

\bibitem[{{Prudnikov} {et~al.}(1986){Prudnikov}, {Brychkov}, \&
  {Marichev}}]{Prudnikov1986}
{Prudnikov}, A.~P., {Brychkov}, Y.~A., \& {Marichev}, O.~I. 1986, {Integrals
  and Series, Vol. 1: Elementary Functions} (Gordon \& Breach, New York)

\bibitem[{{Sato} {et~al.}(2011){Sato}, {Ichiki}, \& {Takeuchi}}]{Sato2011}
{Sato}, M., {Ichiki}, K., \& {Takeuchi}, T.~T. 2011, \prd, 83, 023501

\bibitem[{{Schneider} \& {Hartlap}(2009)}]{Schneider2009}
{Schneider}, P. \& {Hartlap}, J. 2009, \aap, 504, 705

\bibitem[{Seljak \& Bertschinger(1993)}]{Seljak1993}
Seljak, U. \& Bertschinger, E. 1993, Astrophys. J., 417, L9

\bibitem[{{Whittaker} \& {Watson}(1963)}]{Whittaker1963}
{Whittaker}, E.~T. \& {Watson}, G.~N. 1963, {A course of modern analysis}, 4th
  edn. (Cambridge University Press)

\bibitem[{{Wilking} \& {Schneider}(in preparation)}]{Wilking2011}
{Wilking}, P. \& {Schneider}, P. in preparation

\end{thebibliography}

\end{document}